\documentclass[12pt]{article}
\usepackage{amsmath,amssymb}
\newcommand{\beqn}{\begin{eqnarray}}
\newcommand{\eeqn}{\end{eqnarray}}
\newtheorem{theorem}{Theorem}
\newtheorem{proposition}{Proposition}
\newtheorem{lemma}{Lemma}
\newtheorem{corollary}{Corollary}
\newtheorem{defn}{Definition}

\newcommand{\proof}{\noindent {\it Proof.\/}\ }
\newcommand{\qed}{{\it Q.E.D.\/} \bigskip\par}
\newcommand{\remark}{\noindent {\bf Remark.}\ }
\newcommand{\bff}{\pmb{f}}
\newcommand{\bfzero}{\pmb{0}}

\newcommand{\tor}{\mathrm{tor}}
\newcommand{\tr}{\mathrm{tr}}
\newcommand{\hInt}{{\mathbb{Z}}+\frac{1}{2}}
\newcommand{\bphi}{{\phi}^*}
\newcommand{\bpsi}{{\psi}^*}
\newcommand{\half}{\frac{1}{2}}
\newcommand{\defeq}{\stackrel{\mathrm{def}}{=}}
\newcommand{\bfx}{\pmb{x}}
\newcommand{\bfy}{\pmb{y}}
\newcommand{\bfa}{\pmb{a}}
\newcommand{\bfb}{\pmb{b}}
\newcommand{\bfc}{\pmb{c}}
\newcommand{\cbfy}{\check{\pmb{y}}}
\newcommand{\bfalpha}{\pmb{\alpha}}
\newcommand{\bfbeta}{\pmb{\beta}}
\newcommand{\diag}{\mathrm{diag}}
\newcommand{\bbZ}{\mathbb{Z}}
\newcommand{\bbC}{\mathbb{C}}
\newcommand{\frakgl}{\mathfrak{gl}}
\newcommand{\fraksl}{\mathfrak{sl}}
\newcommand{\frakg}{\mathfrak{g}}
\newcommand{\frakd}{\mathfrak{d}}

\newcommand{\id}{\mathrm{id}}

\newcommand{\SL}{\mathop{{\mathrm{SL}}}}

\newcommand{\ord}{\mathop{{\mathrm{ord}}}}
\newcommand{\vac}{{\mathbf{vac}}}

\newcommand{\calF}{{\cal F}}
\newcommand{\calA}{{\cal A}}
\newcommand{\calB}{{\cal B}}
\newcommand{\calD}{{\cal D}}
\newcommand{\calK}{{\cal K}}
\newcommand{\rd}{\partial}
\newcommand{\A}{A} 
\newcommand{\X}{D} 

\begin{document}
\title{
Toroidal Lie algebras and Bogoyavlensky's\\ 
$2+1$-dimensional equation}
\author{
Takeshi Ikeda\\
{\normalsize Department of Applied Mathematics,
Okayama University of Science}\\
{\normalsize Ridaicho 1-1, Okayama 700-0005, Japan}\\
{\normalsize E-mail: ike@xmath.ous.ac.jp}\\
Kanehisa Takasaki\\
{\normalsize Department of Fundamental Sciences, 
Kyoto University}\\
{\normalsize Yoshida, Sakyo-ku, Kyoto 606-8501, Japan}\\
{\normalsize E-mail: takasaki@math.h.kyoto-u.ac.jp}}
\date{}
\maketitle

\begin{abstract}
We introduce an extension of the $\ell$-reduced
KP hierarchy, which we call the $\ell$-Bogoyavlensky hierarchy.
Bogoyavlensky's $2+1$-dimensional extension of the KdV equation
is the lowest equation of the hierarchy in case of $\ell=2$.
We present a group-theoretic characterization of this hierarchy
on the basis of the $2$-toroidal Lie algebra ${\fraksl}_\ell^\tor$.
This reproduces essentially the same Hirota bilinear equations as
those recently introduced by Billig and Iohara et al. We can
further derive these Hirota bilinear equation from a Lax formalism
of the hierarchy.
This Lax formalism also enables us to construct
a family of special solutions that generalize the
breaking soliton solutions of Bogoyavlensky.
These solutions contain the $N$-soliton solutions,
which are usually constructed by use of vertex operators.
\end{abstract}

\renewcommand{\theequation}{\arabic{section}.\arabic{equation}}
\section{Introduction}
\setcounter{equation}{0}
In the early 1980s, Date et al. \cite{bib:DJKM} discovered 
the remarkable fact
that the Lie algebra $\frakgl(\infty)$
acts on the solutions of
the KP(Kadomtsev-Petviashvili) hierarchy. 
The link between infinite-dimensional Lie algebras
and soliton equations has been widely extended.
In particular, Kac and Wakimoto \cite{bib:KW} developed
a scheme to construct hierarchies from vertex operator
representations of general affine Lie algebras.
On these backgrounds, Billig \cite{bib:B2} and 
Iohara et al. \cite{bib:ISW1}
derived new hierarchies of Hirota 
bilinear equations from 
$2$-toroidal Lie algebra $\frakg^\tor$,
a central extension of double loop
algebra $\frakg[s^{\pm 1},t^{\pm 1}]$
of a simple Lie algebra $\frakg$.
Applying the construction 
to the case $\frakg^\tor=\fraksl_\ell^\tor$
(and their principal vertex realization), we 
have an extension of the $\ell$-reduced 
KP hierarchy
of Hirota bilinear form.

For the simplicity of exposition, 
we here deal with the case $\frakg=\fraksl_2$.
The 
typical Hirota bilinear equations of 
lower degrees in hierarchy of 
\cite{bib:B2},\cite{bib:ISW1} are the following 
\beqn
    \left(D_{x}^4-4D_{x}D_{t}\right)\tau\cdot\tau=0,\label{eq:KdV-Hir}\\
    \left(D_{y}D_{x}^3+2D_{y}D_{t}-6D_{z}D_{x}\right)
      \tau\cdot\tau=0\label{eq:Bog-Hir}.
\eeqn
If we count the degrees of Hirota's $D$-operators
$D_y,D_x,D_z,D_t$ as $0,1,2,3$ respectively,
the Hirota bilinear 
equations above have degrees $4$ and $3$ respectively.
The first one (\ref{eq:KdV-Hir}) is a famous bilinear form of 
the KdV equation $4u_t=6uu_x+u_{xxx},$
where $u\defeq 2(\log\tau)_{xx}.$
The second one (\ref{eq:Bog-Hir}) can be written as a
non-linear equation known as 
Bogoyavlensky's $2+1$-dimensional equation \cite{bib:Bog}
\beqn
    u_z=\frac{1}{4}u_{xxy}+vu_x+uu_y,\label{eq:Bog0}
\eeqn
where we set $v\defeq(\log\tau)_{xy}$ and used the KdV equation to 
eliminate the terms including derivative $\rd_t.$
Equation (\ref{eq:Bog0}) has the following 
equivalent Lax form
\beqn
    \frac{\rd P}{\rd z}=[P\rd_y+C,P]\quad\mbox{where}\;
    P\defeq \rd_x^2+u,\;
    C\defeq v\rd_{x}+\frac{3}{4}u_y.\label{eq:Lax}
\eeqn
The main purpose of this paper, 
motivated by the fact above,
is to establish a Lax formalism of
the Hirota bilinear equations
arising from $\fraksl_\ell^\tor$
for any $\ell\ge 2$.
For this purpose, we shall introduce
the $\ell$-{\it Bogoyavlensky hierarchy\/}.
We shall also construct a family of special solutions
that generalize the
{\it breaking solitons\/} studied by Bogoyavlensky.

We shall develop our discussions as follows.
In section \ref{section:rep}, we discuss 
the hierarchy on
a viewpoint of representation theory.
First we give definition of the Lie algebra
$\widetilde{\frakg}^\tor$,
and then explain a general result on their
representations(Lemma \ref{lem:tor_ext}). Using this
lemma, we construct a representation
of $\widetilde{\fraksl}_\ell^\tor$, based on 
the $\ell$-reduction procedure of $\frakgl(\infty)$
in \cite{bib:DJKM}.
We shall show that every vector $\tau$ in the 
$\SL_\ell^\tor$-orbit of the 
\textit{vacuum vector} satisfies 
a hierarchy of Hirota bilinear equations,
which is essentially the same as those derived  
in \cite{bib:B2},\cite{bib:ISW1},
although the generating functions are 
apparently different.
In section \ref{section:Lax}, we first give 
a review on the Lax formalism of the KP hierarchy.
Next we present a heuristic introduction to Bogoyavlensky's 
hierarchy.
These arguments lead us to formulate
the $\ell$-{\it Bogoyavlensky hierarchy\/} for each $\ell\geq2,$
where the so-called Wilson-Sato operator plays 
the significant role. 
In section \ref{section:LaxToBilin},
we give a residue formula for the 
formal Baker-Akhiezer functions of the  
hierarchy by the calculus of pseudo-differential operators.
The formula leads to exactly the same system of Hirota bilinear 
equations derived in section \ref{section:rep}. 
Note that the most part of sections 
\ref{section:Lax},\ref{section:LaxToBilin} 
can be read independently 
from the preceding section.
Section \ref{section:sol} is devoted to investigate 
the special solutions
of the hierarchy, namely the Wronskian solutions and
the $N$-soliton solutions.
As an appendix, we give a table of Hirota equations of 
low degree in section \ref{section:app}.

It should be mentioned that the Bogoyavlensky's equation
is a dimensional reduction, to
$2+1$ dimensions, of the four dimensional
self-dual Yang-Mills equation. Mason and
Sparling \cite{bib:MS} pointed out that
many integrable systems in $2+1$ and 
lower dimensions (in particular, $1+1$
dimensional soliton equations such as
the KdV equation) can be thus derived 
from the self-dual Yang-Mills equation.
In fact, Schiff \cite{bib:Sc} rediscovered
Bogoyavlensky's equation, having been
unaware of Bogoyavlensky's work. Schiff's
work was further extended by Yu et al.
\cite{bib:YTSF,bib:YTF}.

Throughout the paper, the base field is the field $\bbC$
of complex numbers. The symbol $\bbZ$ stands for the set of 
integers. The binomial coefficients $\binom{n}{k}$ is defined as 
$n(n-1)\cdots(n-k+1)/k!$ for $k\geq 0.$
For a vector space $V$, we denote by 
$V[z^{\pm 1}],\,V[z^{\pm 1},w^{\pm 1}]$ 
the spaces $V\otimes \bbC[z,z^{-1}],\,
V\otimes \bbC[z,z^{-1},w,w^{-1}]$ 
of Laurent polynomials with the coefficients in $V$ 
respectively. For an integer $\ell\geq 2,$ we denote
by $\frakgl_\ell$ (resp. $\fraksl_\ell$) the Lie algebras
of the all (resp. traceless) 
$\ell\times\ell$ matrices.

\textbf{Note added:} After the first version of this paper
had written, we learned that F. Calogero \cite{bib:Cal} had studied
equation (\ref{eq:Bog0}) in 1975. We thank Kouichi Toda and his 
collaborators for the information.

\tableofcontents

\section{Deriving hierarchy from representation theory}\label{section:rep}
\setcounter{equation}{0}
\subsection{Definitions of the toroidal Lie algebra}

Let $\frakg$ be a finite-dimensional simple 
Lie algebra over $\bbC.$ 
Let $R$ be the ring of Laurent polynomials of 
two variables $\bbC[s^{\pm 1},t^{\pm 1}].$ 
The module of K\"{a}hler differentials 
$\Omega_R$ of $R$ is defined with the canonical 
derivation $d:R\rightarrow \Omega_R$.
As an $R$-module, $\Omega_R$ is freely generated by 
$ds$ and $dt.$
Let $\overline{\cdot}:\Omega_R\rightarrow \Omega_R/dR$ 
be the canonical projection.
Let $\calK$ denote $\Omega_R/dR.$
Let $(\cdot|\cdot)$ be the normalized Killing form
(\cite{bib:Kacbook}) on $\frakg.$
We define the Lie algebra structure on 
${\frakg}^\tor\defeq{\frakg}\otimes R\oplus \calK$ by
\beqn
   [X\otimes f,Y\otimes g]=[X,Y]\otimes fg+(X|Y)\overline{(df)g},\quad
   [\calK,{\frakg}^\tor]=0\label{eq:bracket-tor}.
\eeqn
In the paper \cite{bib:Kass}, Kassel proved
that the bracket defines a {\textit{universal central 
extension}} of $\frakg\otimes R$ as a Lie algebra
over $\bbC,$ for a large class of commutative
algebra $R.$ 
One can find in \cite{bib:MEY} a simpler proof
(a natural extension of Wilson's discussions \cite{bib:Wil}) of 
the Kassel's result
that works for the cases when the ground field of $R$
has characteristic $0.$

Let $\calD=R{\rd_{\log s}}\oplus R{\rd_{\log t}}$ 
be the Lie algebra of derivations on $R.$
A derivation $\delta\in\calD$ can be 
naturally extended to a derivation on 
the Lie algebra $\frakg\otimes R$ by
\beqn
   \delta\left(X\otimes f\right)\defeq 
   X\otimes\delta f.\label{eq:D-act-on-g}
\eeqn
It is known that a derivation $\delta\in\calD$ acting on 
the Lie algebra $\frakg\otimes R$ has 
a natural extension to $\frakg^\tor$.
The action on the center $\calK$ is given explicitly 
as follows \cite{bib:EM}. 
First we shall define the action of $\calD$
on $\Omega_R$ by:  
\beqn
    f\rd_{\log u}\left({gd\log v}\right)
    \defeq\left(f\rd_{\log u}g\right)
    d\log v+\delta_{u,v}\,{gdf}\quad 
    \mbox{for}\;u,v=s,t.\label{eq:D-act-on-Omega}
\eeqn
The action preserves the exact forms $dR,$ namely we have
$f\rd_{\log u}(dg)=d\left(f\rd_{\log u}g\right)$ for 
$u=s,t,$ 
and so induces an action, also denoted by the same
notation, on $\calK=\Omega_R/dR.$

Let $\check\calD$ denote the Lie subalgebra $R\rd_{\log t}$ of $\calD.$
We shall add the derivations $\check\calD$ to 
$\frakg^\tor$ to get the Lie algebra 
\beqn
    \widetilde{\frakg}^\tor=\frakg^\tor\oplus \check\calD.
    \nonumber
\eeqn
Here the bracket in $\frakg^\tor$ is given by 
(\ref{eq:bracket-tor}).
We define the bracket between $\check\calD$ and $\frakg^\tor$ by the 
action (\ref{eq:D-act-on-g}), and (\ref{eq:D-act-on-Omega}),
$
[\delta,a]=\delta(a)\;(\delta\in \check\calD,\;a\in \frakg^\tor).
$
The bracket in $\check\calD$ is defined by
\beqn
    [f\rd_{\log t},g\rd_{\log t}]
     =\left(f(\rd_{\log t}g)-g(\rd_{\log t}f)\right)\rd_{\log t}
     -\overline{(\rd_{\log t}g)d(\rd_{\log t}f)}.\nonumber
\eeqn
Note that if $f,g\in \bbC[t^{\pm 1}]$ 
in the above formula, then it is 
equivalent to the relation of the Virasoro algebra.
Remark that $\tilde{\frakg}^\tor$ 
is {\textit{not}} the semidirect 
product of $\frakg^\tor$ with the action
of $\check\calD$ by (\ref{eq:D-act-on-g}),(\ref{eq:D-act-on-Omega}).

We have, for $u=s,t$, the Lie subalgebras
\beqn
    \widehat{\frakg}_u\defeq\frakg\otimes{\bbC}[u^{\pm 1}]
    \oplus{\bbC}\,\overline{d\log u},\nonumber
\eeqn
with the brackets given by
\beqn
    [Xu^m,Yu^n]=u^{m+n}[X,Y]+m\delta_{m+n,0}(X|Y)\,
    K_u,\nonumber
\eeqn
which are isomorphic to the affine Lie algebra $\widehat{\frakg}$
with the canonical central element $K_u\defeq\overline{d \log u}.$

We prepare the generating series of $\widetilde{\frakg}^\tor$ as follows:
\beqn
    A_m(z)\defeq\sum_{n\in{\bbZ}}As^nt^m\cdot z^{-n-1},\quad
    D_m^t(z)\defeq\sum_{n\in{\bbZ}}s^nt^m\rd_{\log t}\cdot z^{-n-1},\nonumber\\
    K_m^s(z)\defeq\sum_{n\in \bbZ}\overline{s^nt^md \log s}\cdot z^{-n},\quad
    K_m^t(z)\defeq\sum_{n\in \bbZ}\overline{s^nt^md \log t}\cdot z^{-n-1},\nonumber
\eeqn
where $m\in \bbZ$ and $A\in\frakg.$
The relation $\overline{d(s^nt^m)}=0$ can be neatly expressed by 
these generating series as  
\beqn
    -\frac{\rd}{\rd z}K_m^s(z)+mK_m^t(z)=0.\nonumber
\eeqn

\subsection{A Heisenberg subalgebra $\frakd$ of 
$\widetilde{\frakg}^\tor$}
The Lie algebra $\widetilde{\frakg}^\tor$ has a Heisenberg
subalgebra $\frakd$ with basis  
$\{s^n \rd_{\log t},\overline{s^n d\log t},d\log s\;(n\in\bbZ)\}.$
In fact, if we set $\phi_n\defeq\overline{s^n d\log t},$
$\phi_n^*\defeq s^n \rd_{\log t},$ 
and $\hslash\defeq\overline{d \log s},$ then we have
\beqn
    [\bphi_m,\phi_n]=m\delta_{m+n,0}\hslash,\quad 
    [\phi_m,\phi_n]=[\bphi_m,\bphi_n]=[\hslash,{\frakd}\,]=0.
\nonumber
\eeqn
The Heisenberg algebra $\frakd$ is degenerate in the sense that 
$[\bphi_0,\phi_0]=0.$
We also note that $\frakd$ commute with $\widehat{\frakg}_s.$

Next we realize the action of $\frakd$ on
the space of polynomials
\beqn
    F_{\frakd}\defeq\bbC[y_\ell,y_{2\ell},y_{3\ell},\ldots;
      y^*_\ell,y^*_{2\ell},y^*_{3\ell},\ldots]
      \otimes\bbC[e^{y_0},e^{-y_0}],\quad
    \vac_\frakd\defeq 1\in F_{\frakd}\nonumber
\eeqn
where the action of $\frakd$ is defined by:
\beqn
    \phi_m\mapsto\begin{cases}
      \frac{\rd}{\rd{y^*_{m\ell}}}\quad &m>0\\
      -my_{-m\ell}\quad &m\leq 0
    \end{cases},\quad
    \bphi_m\mapsto\begin{cases}
      \frac{\rd}{\rd{y_{m\ell}}}\quad &m\geq 0\\
      -my^*_{-m\ell}\quad &m<0
    \end{cases},
\quad
\hslash={\mathrm{id}}.\nonumber
\eeqn
For convenience' sake,
we adopted rather peculiar notation for the variables
$y_{n\ell}$ etc. 

We define the generating series 
\beqn
    \phi(z)\defeq\sum_{n\in\bbZ}\phi_nz^{-n-1},\quad
    \bphi(z)\defeq\sum_{n\in\bbZ}\bphi_nz^{-n-1},\nonumber
\eeqn
and for each $m\in\bbZ$, the following {\textit{vertex operators}}
\beqn
    V_m(z)\defeq\exp\left(m\sum_{n>0}y_{n\ell}z^{n}\right)
      e^{my_0}
      \exp
      \left(
        -m\sum_{n>0}\frac{1}{n}\frac{\rd}{\rd{y^*_{n\ell}}}z^{-n}
      \right).\label{eq:vertex-phi}
\eeqn
We also introduce the normal product 
$
:\!\!\phi(z)V_m(z)\!\!:\;\defeq\phi(z)_{<0}V_m(z)+V_m(z)\phi(z)_{>0}\nonumber
$
where $\phi(z)_{<0}\defeq\sum_{n<0}\phi_n z^{-n-1}$
and $\phi(z)_{>0}\defeq\sum_{n>0}\phi_n z^{-n-1}.$

We shall say a representation $V$ of the affine Lie algebra
$\widehat{\frakg}_s$ is 
{\textit{restricted}} 
if and only if for all $v\in V$ and $X\in\frakg$,
there exists $N$ such that $Xs^n\cdot v=0$ for $\forall n>N.$ 
Now we can state the following lemma due to 
Iohara et al. \cite{bib:ISW1} and Berman-Billig
\cite{bib:BB}.
\begin{lemma}\label{lem:tor_ext}
Let $(V,\pi)$ be a restricted representation of $ \widehat\frakg_s$ 
such that
$\overline{d\log s}\mapsto{\id}_V.$
Then we can define the representation $\pi^\tor$ of $\;\tilde{\frakg}^\tor$ on
$V\otimes \calF_{\frakd}$ such that 
\beqn
    A_m(z)
    &\mapsto A^\pi(z)V_m(z),\quad
    K_m^s(z)
    &\mapsto V_m(z),\nonumber\\
    D_m^t(z)
    &\mapsto\phi^*\!(z)V_m(z),\quad
    K_m^t(z)
    &\mapsto\;:\!\phi(z)V_m(z)\!:\;,\nonumber
\eeqn
where $A\in \frakg,\; m\in \bbZ$ and $A^\pi(z)\defeq\sum_{n\in \bbZ}\pi(As^n)z^{-n-1}.$
\end{lemma}

\subsection{Review on a representation of $\frakgl(\infty)$ and its
$\ell$-reduction}
In the subsection, we collect basic results on 
representation of $\frakgl(\infty)$ which
is used in the sequel. One can find in the original paper \cite{bib:DJKM}
the proofs we omit.
We shall mainly follow the notation used in book \cite{bib:JMD}.

Let the associative $\bbC$-algebra $\calA$ be generated by 
$\psi_i,\bpsi_i(i\in\hInt)$ with the relations:
\beqn
    \psi_i\psi_j+\psi_j\psi_i=\bpsi_i\bpsi_j+\bpsi_j\bpsi_i=0,\quad
    \bpsi_i\psi_j+\psi_j\bpsi_i=\delta_{i+j,0}.\nonumber
\eeqn
Consider a left $\calA$-module with a cyclic vector $\vac_\calA$ satisfying
\beqn
    \psi_i\vac_\calA=\bpsi_i\vac_\calA=0\quad \mbox{for}\quad i>0.\label{eq:vac-psi}
\eeqn 
This $\calA$-module $\calA \vac_\calA$ is called 
the fermionic Fock space,
which we denote by $\calF_\calA.$

We define the following infinite formal Laurent series of 
the variable $\lambda$
\beqn
    \psi(\lambda)\defeq\sum_{i\in\hInt}\psi_i \lambda^{-i-\half},\quad
    \bpsi\!(\lambda)\defeq\sum_{i\in\hInt}\bpsi_i \lambda^{-i-\half},\nonumber
\eeqn
and the normal ordering of the
following quadratic expression to be
\beqn
    :\psi_i\bpsi_j\defeq
    \begin{cases}
    \psi_i\bpsi_j\quad &\mbox{if}\quad i<0\quad\mbox{or}\quad j>0\\
    -\bpsi_j\psi_i\quad&\mbox{if}\quad i>0\quad \mbox{or}\quad j<0
    \end{cases}.\nonumber
\eeqn

Consider the following set of infinite complex matrices
\beqn
\left.\left\{A=(a_{ij})_{i,j\in {\bbZ}+\half}\;\right|\;
a_{ij}=0 \;\mbox{for}\;|i-j|\gg 0
\right\}.\nonumber
\eeqn
It forms a Lie algebra over ${\bbC}$
with the usual bracket $[A,B]_0\defeq AB-BA$,
which we shall denote by $\overline{\frakgl}(\infty)$.
We shall define the 2-cocycle $\omega$ on $\overline{\frakgl}(\infty)$ by :
\beqn
    \omega({E}_{ij},{E}_{i'j'})\defeq\delta_{ji'}\delta_{ij'}
    (\theta(j)-\theta(i))\nonumber
\eeqn
where $E_{ij}\defeq\left(\delta_{ii'}\delta_{jj'}\right)_{i',j'\in \hInt}$ 
is the matrix unit
and $\theta$ is defined by
$\theta(i)\defeq
1 $ if $i> 0$ and $\theta(i)\defeq
0 $ if $i<0.$
Let us introduce the central extension $\frakgl(\infty)\defeq\overline{\frakgl}(\infty)
\oplus{\bbC}c$ with the central element $c$ and the bracket:
\beqn
    [A,B]\defeq[A,B]_0+\omega(A,B)c,\quad A,B\in\overline{\frakgl}(\infty).\nonumber
\eeqn

\begin{lemma}\cite{bib:DJKM}
The map
$
E_{ij}\mapsto :\psi_{-i}\bpsi_j:,c\mapsto {\mathrm{id}_{\calF_\calA}}$ 
defines the representation
$\pi_\infty$ of $\frakgl(\infty)$ on the fermionic Fock space $\calF_\calA.$ 
\end{lemma}
In terms of the
generating series defined by
\beqn
    E(\lambda,\mu)\defeq\sum_{i,j\in\bbZ+\half}
    E_{ij}\lambda^{i-\half}\mu^{-j-\half},\nonumber
\eeqn
the lemma has the equivalent expression
\beqn
    E(\lambda,\mu)
    \mapsto :\!\psi(\lambda)\bpsi\!(\mu)\!\!:\;.\nonumber
\eeqn

Fix an integer $\ell\geq 2.$ 
Let $A$ be a matrix $(a_{ij})_{i,j\in {\bbZ}+\half}$ in $\frakgl(\infty).$
We assume the condition:
\beqn
    a_{i,j}=a_{i+\ell,j+\ell}\quad\mbox{for all}\;
    i,j\in\bbZ+\textstyle\half.\label{eq:l-periodic}
\eeqn
Then we introduce, for each $n\in\bbZ$, the $\ell\times\ell$ 
complex matrix by
\beqn
    A_n\defeq\left(a_{i+\half,j+\half+n\ell}
    \right)_{0\leq i,j\leq\ell-1}\nonumber
\eeqn
and define an $\ell\times\ell$ matrix of Laurent {\textit{polynomial}} of the variable $s$
\beqn
    \sum_{n\in\bbZ}A_ns^n.\nonumber
\eeqn

Conversely, let $\sum_{n}A_n s^n$ be a Laurent polynomial in 
$\frakgl_\ell[s^{\pm 1}].$
Then we associate the 
$\ell$-periodic infinite matrix in the following block form 
\beqn
    \sum_{n}A_ns^n
    \mapsto
        \begin{pmatrix}
        \ddots&\ddots&\ddots&\ddots&\ddots\\
        \ddots&A_0&A_1&A_2&\ddots\\
        \ddots&A_{-1}&A_0&A_1&\ddots\\
        \ddots&A_{-2}&A_{-1}&A_0&\ddots\\
        \ddots&\ddots&\ddots&\ddots&\ddots
        \end{pmatrix}
    \in \frakgl(\infty).\nonumber
\eeqn
Let us denote by $\iota_\ell$
the above map, which is clearly injective.
Let $X,Y\in \frakgl_\ell$ and consider 
$Xs^m,Ys^n$ $(m,n\in\bbZ)$ as elements
of $\frakgl(\infty).$
Then the commutation relation is given by
\beqn
    [Xs^m,Ys^n]=[X,Y]s^{m+n}+m\delta_{m+n,0}\tr(XY)c.\nonumber
\eeqn
So, in particular, if we extend $\iota_\ell$ to 
$\widehat{\fraksl}_\ell=\fraksl_\ell[s^{\pm 1}]\oplus \bbC K_s$
by $\iota_\ell(K_s)\defeq c$, then it gives the embedding 
$\iota_\ell$ of Lie algebras
$\widehat{\fraksl}_\ell$
into $\frakgl(\infty).$

We shall describe a base of $\iota_\ell(\widehat{\fraksl}_\ell)
\cong\widehat{\fraksl}_\ell$
for the later use.
\begin{defn}
For each $\ell$-th root of unity $\zeta\ne 1$, we define
$E_n^\zeta\in \frakgl(\infty)\;(n\in \bbZ)$ by the following series   
\beqn
    E(\lambda,\zeta\lambda)
    =\sum_{n\in \bbZ}E_n^\zeta \lambda^{-n-1}.\nonumber
\eeqn
\end{defn}
It is easy to see that $E_n^\zeta$ satisfies 
$\ell$-periodic condition (\ref{eq:l-periodic}).
If we shall regard $E_n^\zeta$ as an element of 
$\frakgl_\ell[s^{\pm 1}],$
then $E_n^\zeta$ is traceless.
Note that $E_n^\zeta$ is homogeneous of the principal degree $n$, 
here by the principal 
degree of the matrix $s^n e_{ij}$ in
$\frakgl_\ell$ we mean $j-i+n\ell\in\bbZ.$

\bigskip
\textbf{Example 1}
For $\ell=2,\;\zeta=-1$, we have
\beqn
    E^\zeta_{2i+1}
    =\begin{pmatrix}
        0         &  s^i\\
        -s^{i+1}  &    0
    \end{pmatrix},\quad
    E^\zeta_{2i}
    =\begin{pmatrix}
        -s^i      &    0\\
        0         &  s^i
    \end{pmatrix}
    \quad(i\in \bbZ).\nonumber
\eeqn

If we define
$\Lambda_n\defeq\sum_{i\in\bbZ+\half}E_{i,i+n}(n\in\bbZ\setminus \{0\}),$
then we have the following Heisenberg relations
\beqn
    [\Lambda_m,\Lambda_n]=m\delta_{m+n,0}c.\nonumber
\eeqn

\begin{proposition}The matrices
$E_n^\zeta\;(\zeta^\ell=1,\,\zeta\ne 0,\;n\in\bbZ),\;
\Lambda_n\;(n\in\bbZ,\;n\not\equiv 0 \mod\ell)$ form a 
basis of $\fraksl_\ell[s^{\pm 1}].$ 
Therefore the representation 
$\pi_\infty\circ\iota_\ell$ of $\widehat{\fraksl}_\ell$ 
is given by 
\beqn
    \sum_{n\in\bbZ}E_n^\zeta\lambda^{-n-1}
    \mapsto:\psi(\lambda)\bpsi\!(\zeta\lambda):
    \quad (\zeta^\ell=1,\zeta\ne 1),\label{eq:Ezeta_rep}\\
    \Lambda_n\mapsto \sum_{i\in\hInt}:\psi_{-i}\bpsi_{i+n}:
        \;(n\in\bbZ,\;n\not\equiv 0 \mod\ell),\quad
    K_s\mapsto \id.\nonumber
\eeqn
\end{proposition}

\begin{proposition}
If we identify $E^\zeta_m,\Lambda_m$ with the 
elements of ${\frakgl}_\ell[s^{\pm 1}],$
then we have
\beqn
    s^mE^\zeta_n=E^\zeta_{n+m\ell},\quad
    s^m \Lambda_n=\Lambda_{n+m\ell}.\label{eq:mult_s}
\eeqn
\end{proposition}
Note that the centralizer
of $\Lambda_\ell$ in $\frakgl(\infty)$
coincide with $\iota_\ell(\widehat{\fraksl}_\ell)+\bbC 1.$

\begin{lemma}\label{lem:Omega-af}
Let $\Omega_\calA(\lambda)\defeq\psi(\lambda)\otimes \bpsi(\lambda)$ be the operator
on $\calF_{\calA}^\tor\defeq
\calF_\calA\otimes\calF_\calA.$ Then we have
\beqn
    \left[\oint_{\lambda=\infty}
    \lambda^{n\ell}\Omega_\calA(\lambda)d\lambda,\;
    \widehat{\fraksl}_\ell\right]=0
    \quad\mbox{for}\quad \forall n\in \bbZ.\nonumber
\eeqn
\end{lemma}
\proof
We use the expression
\beqn
    \oint_{\lambda=\infty}\lambda^{n\ell}\Omega_\calA(\lambda)d\lambda
    =\sum_{i\in\bbZ+\half}\psi_{-i}\otimes\bpsi_{i+n\ell}.
    \label{eq:Omega-af-psi}
\eeqn
For
$
A=\sum_{i,j\in\hInt}
a_{ij}:\!\!\psi_{-i}\bpsi_j\!\!:
$
that satisfy (\ref{eq:l-periodic}), we have
\begin{align}
    &[A\otimes 1+1\otimes A,\sum_{k}\psi_{-k}\otimes\bpsi_{k+n\ell}]\nonumber\\
    =&\sum_{k}[A,\psi_{-k}]\otimes\bpsi_{k+n\ell}
    +\sum_{k}\psi_{-k}\otimes[A,\bpsi_{k+n\ell}]\nonumber\\
    =&\sum_{k}\sum_{j}a_{jk}\psi_{-j}\otimes\bpsi_{k+n\ell}
    +\sum_{k}\psi_{-k}\otimes\sum_{j}(-a_{k+n\ell,j})\bpsi_{j}\nonumber\\
    =&\sum_{k,j}\left(a_{j,k-n\ell}-a_{j+n\ell,k}\right)\psi_{-j}\otimes\bpsi_{k}=0\nonumber,
\end{align}
where we used (\ref{eq:l-periodic}) in the last line.
\qed

\subsection{Equation for the $\SL_\ell^\tor$-orbit of the vacuum}
Combining the results in the preceding subsection and 
Lemma \ref{lem:tor_ext}, we have the representation
of $\widetilde{\fraksl}_\ell^\tor$ on the space 
${\calF}_{\calA}^\tor\defeq\calF_{\calA}\otimes\calF_{\frakd}$
defined as
$\pi_{\ell,\calA}^\tor\defeq(\iota_\ell\circ\pi_\infty)^\tor.$
Now we are  in a position to define the following 
important operator:
\beqn
\Omega_\calA^\tor(\lambda)\defeq\sum_{m\in\bbZ}
\psi(\lambda)V_m(\lambda^\ell)\otimes 
\bpsi\!(\lambda)V_{-m}(\lambda^\ell).\nonumber
\eeqn
The operator satisfy the following property.
\begin{lemma}
We have
\beqn
\left[\oint_{\lambda=\infty}
\lambda^{j\ell}
\Omega_\calA^\tor(\lambda)d\lambda,\fraksl_\ell^\tor\right]=0
\quad \mbox{for any} \quad j\in \bbZ.\nonumber
\eeqn
Here the contour integral is understood symbolically, 
namely, just to extract the coefficient of $\lambda^{-1}: 
\oint \lambda^n d\lambda/(2\pi i) = \delta_{n,-1}$. 
\end{lemma}
\proof
We first note that the operator $\Omega_{\calA}^\tor(\lambda)$ 
is the product
of $\Omega_{\calA}(\lambda)=\psi(\lambda)\otimes\bpsi(\lambda)$
and $\sum_{m\in\bbZ}
V_m(\lambda^\ell)\otimes V_{-m}(\lambda^\ell)$
that commute with each other.
In view of lemma \ref{lem:Omega-af} and the fact that 
the latter operator is a series of $\lambda^\ell,$
one can see the lemma above is valid.
\qed
Let $\SL_\ell^\tor$ denote a group of invertible linear 
transformations on $\calF_\calA^\tor$ generated
by the exponential action of the elements in  
$\fraksl_\ell\otimes R$ acting locally nilpotently.
\begin{corollary}
Let $\tau\in \calF_\calA^\tor$ be in the 
$\SL_\ell^\tor$-orbit of $\vac_\calA^\tor\defeq
\vac_\calA\otimes\vac_\frakd.$
Then we have 
\beqn
\oint_{\lambda=\infty}
\lambda^{j\ell}
\Omega_\calA^\tor(\lambda)\left(\tau\otimes \tau\right)d\lambda=0\quad 
\mbox{for} \;\forall j\geq 0.\label{eq:Omega_tau}
\eeqn
\end{corollary}
\proof
By the preceding Lemma, it suffices to show
the equations above for $\tau=\vac_\calA^\tor.$
In fact, by the expressions (\ref{eq:Omega-af-psi}) and 
(\ref{eq:vertex-phi}),
one can directly verify this.
\qed

\subsection{Bosonization}
Here we present a summary of 
the {\textit{boson-fermion correspondence}}, that 
says 
the space $\calF_\calA$ can be 
identified with the space of polynomials
\beqn
\calF_\calB\defeq\bbC[x_1,x_2,x_3,\ldots]\otimes 
\bbC[e^{x_0},e^{-x_0}],\quad
\vac_{\calB}\defeq 1\in \calF_\calB.\nonumber
\eeqn
To state the correspondence precisely,
we introduce the {\textit{charged vacua}}
\beqn
\vac_{\calA}^{[n]}
\defeq\begin{cases}
\bpsi_{n+\half}\cdots\bpsi_{-\half}\vac_{\calA}\quad &n<0\\
\vac_{\calA}\quad &n=0\\
\psi_{-n+\half}\cdots\psi_{-\half}\vac_{\calA}\quad &n>0
\end{cases}.\nonumber
\eeqn
\begin{lemma}\cite{bib:DJKM} There exists an unique linear 
isomorphism
$
\sigma:\calF_\calA\rightarrow\calF_\calB
$
such that
\beqn
\sigma(\vac_{\calA}^{[n]})=\vac_{\calB}^{[n]},
\quad
\sigma\Lambda_m\sigma^{-1}=\begin{cases}\frac{\rd}{\rd x_m}&
\quad m>0\\
-mx_{-m}&\quad m<0\end{cases},\nonumber
\eeqn 
where we set $\vac_{\calB}^{[n]}\defeq e^{nx_0}\;(n\in\bbZ).$
\end{lemma}

We introduce {\textit {the vertex operators}} by
\beqn
X(\lambda)&\defeq&
\exp
\Bigl(
\xi(\bfx,\lambda)
\Bigr)
e^{x_0}\lambda^{\rd_{x_0}}
\exp
\Bigl(
-\xi(\widetilde{\rd}_{\bfx},\lambda^{-1})
\Bigr),\nonumber\\
X^*(\lambda)&\defeq&
\exp
\Bigl(
-\xi(\bfx,\lambda)
\Bigr)
e^{-x_0}\lambda^{-\rd_{x_0}}
\exp
\Bigl(
\xi(\widetilde{\rd}_{\bfx},\lambda^{-1})
\Bigr),\nonumber\\
X(\lambda,\mu)&\defeq&
\exp\Bigl(\xi(\bfx,\lambda)-\xi(\bfx,\mu)\Bigr)
\lambda^{\rd_{x_0}}\mu^{-\rd_{x_0}}
\exp\Bigl(-\xi(\widetilde{\rd}_{\bfx},\lambda^{-1})
+\xi(\widetilde{\rd}_{\bfx},\mu^{-1})
\Bigr)\nonumber
\eeqn
where $\widetilde{\rd}_{\bfx}$ stands for 
$(\rd_{x_1},\frac{1}{2}\rd_{x_2},\frac{1}{3}\rd_{x_3},\ldots)$
and we set
\beqn
\xi(\bfx,\lambda)\defeq\sum_{n=1}^\infty
x_n\lambda^n\label{eq:def_xi}
\eeqn
Then we have the following correspondences of operators:
\beqn
&\sigma\psi(\lambda)\sigma^{-1}=X(\lambda),\quad
\sigma\bpsi(\lambda)\sigma^{-1}=X^*\!(\lambda),\nonumber\\
&\sigma\Bigl(
:\psi(\lambda)\bpsi(\mu):
\Bigr)\sigma^{-1}
=\displaystyle\frac{1}{\lambda-\mu}
\Bigl(
X(\lambda,\mu)-1
\Bigr),\nonumber
\eeqn
where
$
\frac{1}{\lambda-\mu}
$
stands for the series
$
\sum_{n=0}^\infty\lambda^{-n-1}\mu^n.
$

\subsection{Hirota bilinear equation arising from toroidal Lie algebras}
If we  translate equation (\ref{eq:Omega_tau}) into bosonic language,
then it comes out  a hierarchy of Hirota bilinear equations.
So let us introduce the following bosonic counterpart of operator 
$\Omega_\calA^\tor(\lambda)$
\beqn
\Omega_\calB^\tor(\lambda)\defeq
\sum_{m\in {\bbZ}}X(\lambda)V_m(\lambda^\ell)\otimes
X^*(\lambda)V_{-m}(\lambda^\ell)
=(\sigma\otimes \id)\Omega_\calA^\tor(\lambda)(\sigma^{-1}\otimes\id).
\nonumber
\eeqn
We define $\calF_\calB^\tor\defeq\calF_\calB\otimes\calF_\frakd$
and $\vac_\calB^\tor\defeq\vac_\calB\otimes\vac_\frakd$.
To state next theorem, we define the elementary Schur 
polynomials $p_n(\bfx)\;(n=0,1,\ldots)$ by the
generating series 
\beqn
\sum_{n=0}^\infty p_n(\bfx)\lambda^n
=\exp\left(\sum_{k=1}^\infty x_k \lambda^k \right),
\nonumber
\eeqn
here $\bfx$ stands for $(x_1,x_2,x_3,\ldots).$

\textbf{Notation}
We shall use the notation $\cbfy\defeq(y_\ell,y_{2\ell},\ldots)$,
$\bfy^*\defeq(y^*_\ell,y^*_{2\ell},\ldots)$, $y\defeq y_0$, 
and $\bfy=(y,\cbfy)$ in the sequel of this paper.

\begin{theorem}
Let $\tau=\tau(\bfx,\bfy)$ be in the 
$\SL_\ell^\tor$-orbit of 
$\vac_{\calB}^\tor
\in \calF_\calB^\tor.$
Then $\tau$ satisfies the bilinear equation:
\beqn
\sum_{m,k=0}^\infty
p_m(2\bfa)p_{m+k\ell+j\ell+1}(-\widetilde{D_{\bfx}})p_{k}((\kappa-D_y)\bfb)
e^{\langle \bfa,D_{\bfx}\rangle+\langle \bfb,D_{\cbfy}\rangle}
\tau(\bfx,\bfy)\cdot\tau(\bfx,\bfy)=0\nonumber
\eeqn
for $j\geq 0$
where $\bfa=(a_1,a_2,...),\;\bfb=(b_\ell,b_{2\ell},\ldots)$ 
and 
$\kappa$ are indeterminates and
\beqn
\widetilde{D_{\bfx}}\defeq(D_{x_1},\frac{D_{x_2}}{2},\frac{D_{x_3}}{3},\ldots),\;
\langle \bfa,D_{\bfx}\rangle\defeq
\sum_{n=1}^\infty a_n D_{x_n},
\;\langle \bfb,D_{\cbfy}\rangle\defeq
\sum_{n=1}^\infty b_{n\ell} D_{y_{n\ell}}.\label{eq:BogHirota}
\eeqn
\end{theorem}

\proof
By the bosonic realization, we shall identify
${\calF}_{\calB}^\tor\otimes{\calF}_{\calB}^\tor$ 
with the following space of polynomials
\beqn
\bbC[e^{\pm x_0^{(i)}},e^{\pm y_0^{(i)}},
x_n^{(i)},y_{n\ell}^{(i)},{y_{n\ell}^{*(i)}}\;
(i=1,2,\,n=1,2,\ldots)].
\nonumber
\eeqn
To convert equation (\ref{eq:Omega_tau}) into
bosonic language, 
we shall introduce new variables,
\begin{alignat}{3}
    x_n&\defeq\half(x_n^{(1)}-x_n^{(2)}),
    &\qquad
    \overline{x}_n
    &\defeq\half(x_n^{(1)}+x_n^{(2)})
    &\quad 
    \mbox{for}\quad n\ge 0,\nonumber\\
    y_{n\ell}&\defeq\half(y_{n\ell}^{(1)}-y_{n\ell}^{(2)}),
    &\qquad
    \overline{y_{n\ell}}
    &\defeq\half(y_{n\ell}^{(1)}+y_{n\ell}^{(2)})
    &\quad 
    \mbox{for}\quad n\ge 0,\nonumber\\
    y_{n\ell}^{*}&\defeq\half(y_{n\ell}^{*(1)}-y_{n\ell}^{*(2)}),
    &\qquad
    \overline{{y_{n\ell}^*}}
    &\defeq
    \half(y_{n\ell}^{*(1)}+y_{n\ell}^{*(2)})    &\quad 
    \mbox{for}\quad n\ge 1.\nonumber
\end{alignat}
Then we obtain
\beqn
X(\lambda)\otimes X^*\!(\lambda)
=\exp\left(2\sum_{n=1}^\infty x_n{\lambda^{n}}
\right)
e^{2x_0}\lambda^{\rd_{x_0}}
\exp\left(
-\sum_{n=1}^\infty\frac{1}{n}\frac{\rd}{\rd x_n}\lambda^{-n}
\right)\nonumber
\eeqn
and
\beqn
&\displaystyle\sum_{m\in\bbZ}&V_m(\lambda^\ell)
\otimes V_{-m}(\lambda^\ell)\nonumber\\
=
&\displaystyle\sum_{m\in\bbZ}&
\exp\left(
2m\sum_{n=1}^\infty
y_{n\ell}\lambda^{n\ell}
\right)e^{2my_0}
\exp\left(
-m\sum_{n=1}^\infty
\frac{1}{n}\frac{\rd}{\rd y^{*}_{n\ell}}\lambda^{-n\ell}
\right).\nonumber
\eeqn
Using a lemma by Billig (\cite{bib:B2}, Proposition 3, 
see also \cite{bib:ISW2}),
we obtain the Hirota bilinear equation.
Here we made a reduction
$D_{y^*_{n\ell}}=0,$ since,
by the construction of the representation, $\tau$
in the orbit does not depend on the variables
$y^*_{n\ell}.$
\qed

\remark
\begin{enumerate}
\item By the construction of the representation, 
it is clear that our $\tau=\tau(\bfx,\bfy)$
does not depend on the variables
$x_\ell,x_{2\ell},\ldots.$

\item If we put $b_\ell=b_{2\ell}=\cdots=0$,
then equation (\ref{eq:BogHirota}) is reduced to 
\beqn
\sum_{m=0}^\infty
p_m(2\bfa)p_{m+j\ell+1}(-\widetilde{D_{\bfx}})
e^{\langle \bfa,D_{\bfx}\rangle}
\tau\cdot\tau=0\quad (j\geq 0)\nonumber
\eeqn
This is known as 
bilinear equation for the $\ell$-reduced KP hierarchy \cite{bib:DJKM}.
\end{enumerate}

\bigskip
\textbf{Example 2}
Let $\ell=2$.
From the coefficient of $a_3$ for $j=0$ , we have a equation:
\beqn
\left(
D_{x}^4-4D_xD_t
\right)\tau\cdot\tau=0,\quad \mbox{where}\quad x=x_1,t=x_3.
\eeqn
This can be transformed into a non-linear
equation
\beqn
2\rho_{xt}=\frac{1}{2}\rho_{xxxx}+3\rho_{xx}^2
\eeqn
where we set
$\rho=\log\tau.$
Differentiating by $x$, we have the original KdV equation:
\beqn
u_t=\frac{1}{4}u_{xxx}+\frac{3}{2}uu_x, \quad u\defeq 2\rho_{xx}.
\label{eq:KdV1}
\eeqn

\bigskip
\textbf{Example 3}
Let $\ell=2$. 
The coefficient of $b_2$ in (\ref{eq:BogHirota}) for $j=0$ gives an equation
\beqn
\left(
\frac{1}{6}D_{y_0}D_{x_1}^3
+\frac{1}{3}D_{y_0}D_{x_3}
-D_{x_1}D_{y_2}
\right)
\tau\cdot\tau=0.\label{eq:BogHir}
\eeqn
Equation (\ref{eq:BogHir}) 
times $6$ is nothing but (\ref{eq:Bog0})
with $y_0=y,x_1=x,y_2=z,x_3=t.$

We shall use the following formulae of 
the logarithmic transformations:
\beqn
\frac{{D_x}^3{D_y}\tau\cdot \tau}{\tau^2}
=12\rho_{xx}\rho_{xy}+2\rho_{xxxy},\quad
\frac{{D_x}{D_y}\tau\cdot \tau}{\tau^2}
=2\rho_{xy},
\eeqn
where $\rho\defeq\log\tau.$
These allows us to write equation (\ref{eq:BogHir}) as
a non-linear equation:
\beqn
2\rho_{xy_2}=
2\rho_{xx}\rho_{xy}+\frac{1}{3}\rho_{xxxy}
+\frac{2}{3}\rho_{yt}.\label{eq:rhoBog}
\eeqn

\bigskip
\textbf{Example 4} 
Let us derive Bogoyavlensky's $2+1$-dimensional equation
by our hierarchy of Hirota equations.
We differentiate (\ref{eq:rhoBog}) by $x$ to give:
\beqn
2\rho_{xxy_2}&=&2(\rho_{xx}\rho_{xy})_{x}
+\frac{1}{3}\rho_{xxxxy}+\frac{2}{3}\rho_{xyt}\\
&=&2\rho_{xxx}\rho_{xy}+2\rho_{xx}\rho_{xxy}
+\frac{1}{3}\rho_{xxxxy}
+\frac{1}{3}\left(\frac{1}{2}\rho_{xxxx}+3\rho_{xx}^2
\right)_y\\
&=&\frac{1}{2}\rho_{xxxxy}+2\rho_{xxx}\rho_{xy}
+4\rho_{xx}\rho_{xxy}.
\eeqn
We have used equation (\ref{eq:KdV1})  
in the second line. 
Putting $u=2\rho_{xx}$, it reads
\beqn
u_{y_2}=\frac{1}{4}u_{xxy}
+\frac{1}{2}(\rd_x^{-1}u_y)u_x+uu_y,\label{eq:Bog2}
\eeqn
where we interpret $\rd_x^{-1}u_y$ as $2\rho_{xy}.$ 
After the simple scale transformation 
$\rd_x=2\rd_{x'},\rd_{y_2}=-\rd_{y'_2},u=-4u'$,
(\ref{eq:Bog2}) coincide with (1.22) in Chap. II in \cite{bib:Bog}.

\bigskip
\textbf{Example 5}
Let $\ell=3$.
There is a non-trivial equation of degree
$4$ unique up to scaler,
which is not in the $3$-reduced KP hierarchy
(the Boussinesq hierarchy), 
i.e.
\beqn
\left(
4D_{x_1}D_{y_3}-D_{y_0}D_{x_4}-D_{y_0}D_{x_1}^2D_{x_2}
\right)
\tau\cdot\tau=0.\label{eq:BogHir3}
\eeqn
For instance, this arises from the coefficient 
of $b_3$ in (\ref{eq:BogHirota}) for 
$j=0$. Introducing 
$\rho=\log\tau$, we can rewrite (\ref{eq:BogHir3}) as a non-linear equation:
\beqn
4\rho_{x_1y_3}
-\rho_{y_0x_4}
-\rho_{y_0x_1x_1x_2}
-4\rho_{y_0x_1}\rho_{x_1x_2}
-2\rho_{y_0x_2}\rho_{x_1x_1}=0.
\eeqn

\section{Lax formalism of Bogoyavlensky hierarchy}\label{section:Lax}
\setcounter{equation}{0}

\subsection{Formal pseudo-differential operators}

The Lax formalism of the KP hierarchy is described 
in the language of formal pseudo-differential 
operators (PsDO for short) on a line.  
An affine coordinate $x$ of this line is to be 
identified with the first variable $x_1$ in the 
bosonization of free fermions in the previous section.  
Let $\rd_x$ denote the derivation $\rd/\rd x$.  
In an abstract setting, one can take an arbitrary ring 
with a derivation $\rd$ (namely, a {\it differential ring}); 
we shall rather naively consider, e.g.,  the ring of 
formal power series of $x$.  

A formal PsDO is a formal linear combination, 
$A = \sum_n a_n \rd_x^n$, of integer powers of $\rd_x$ 
with coefficients $a_n = a_n(x)$ that depend on $x$.  
The index $n$ ranges over all integers with an upper 
bound $N$.  The least upper bound is called the 
{\it order} of this PsDO.  The first non-vanishing 
coefficient $a_N$ is called the {\it leading 
coefficient}.  If the leading coefficient is equal 
to $1$, the PsDO is said to be {\it monic}.  
It is convenient to use the following notation: 
\beqn
    (A)_{\ge k} &\defeq& \sum_{n \ge k} a_n \rd_x^n, 
    \nonumber \\
    (A)_{\le k} &\defeq& \sum_{n \le k} a_n \rd_x^n, 
    \nonumber \\
    (A)_k &\defeq& a_k. \nonumber
\eeqn

Addition and multiplication (or composition) of 
two PsDO's are defined as follows.  Addition of 
two PsDO's is an obvious operation, namely, the 
term wise sum of the coefficients.  Multiplication 
is defined by extrapolating the Leibniz rule 
\beqn
    \rd_x^n \circ f 
    = \sum_{k \ge 0} \binom{n}{k} f^{(k)} \rd_x^{n - k},\nonumber 
\eeqn
to the case where $n$ is negative.  Here ``$\circ$'' 
stands for composition of two operators, and 
$f^{(k)}$ the $k$-th derivative $\rd^k f/\rd x^k$ 
of $f$.  More explicitly, the product $C = A \circ B$ 
of two PsDO's $A = \sum_n a_n \rd_x^n$ and 
$B = \sum_n b_n \rd_x^n$ is given by 
\beqn
    C = \sum_{m,n,k} \binom{m}{k} a_m^{(k)} b_n \rd_x^{m+n-k}.\nonumber 
\eeqn
Note that the $n$-th order coefficient $c_n = (C)_n$ 
is the sum of a finite number of terms.  We shall 
frequently write $AB$ rather than $A \circ B$ if 
it does not cause confusion.  

Another important operation is the formal adjoint 
$A \mapsto A^*$: 
\beqn
    A = \sum_n a_n \rd_x^n \longmapsto 
    A^* = \sum_n (-\rd_x)^n \circ a_n.\nonumber 
\eeqn
This is an anti-homomorphism, namely, for any pair 
$A,B$ of PsDO's, 
\beqn
    (AB)^* = B^* A^*. \nonumber
\eeqn

Any PsDO $A = \sum_{n \le N} a_n \rd_x^n$ with 
an invertible leading coefficient $a_N$ has an 
inverse PsDO.  In particular, a monic PsDO is 
invertible.

\subsection{KP hierarchy and its $\ell$-reductions}

The standard Lax formalism of the KP hierarchy 
uses a monic first order PsDO (the {\it Lax 
operator}) of the form 
\beqn
    L \defeq \rd_x + \sum_{n=1}^\infty g_{n+1}\rd_x^{-n}.\nonumber 
\eeqn
The hierarchy is defined by the Lax equations 
\beqn
    \frac{\rd L}{\rd x_n} = [B_n, L]\nonumber
\eeqn
with an infinite number of {\it time variables} 
$\bfx = (x_1,x_2,\ldots)$.  The {\it Zakharov-Shabat 
operators} $B_n$ are given by 
\beqn
    B_n \defeq ( L^n )_{\ge 0}. \nonumber
\eeqn
We identify $x_1$ with $x$, namely $x_1 = x$. This is 
consistent with the first Lax equation 
\beqn
    \frac{\rd \Psi(\lambda)}{\rd x_1} 
    = [\rd_x, \Psi(\lambda)] 
    = \frac{\rd \Psi(\lambda)}{\rd x}.\nonumber 
\eeqn

These Lax equations are associated with the linear 
equations 
\beqn
    L \Psi(\lambda) = \lambda \Psi(\lambda), \quad 
    \frac{\rd \Psi(\lambda)}{\rd x_n} = B_n \Psi(\lambda). \nonumber
\eeqn
Here $\Psi(\lambda)$ (the ``wave function'') is 
understood to be a function of both $\bfx$ and $\lambda$, 
$\Psi(\lambda) = \Psi(\bfx,\lambda)$, though we shall 
frequently omit writing $\bfx$ explicitly.  The Lax 
equations ensure the existence of a non-vanishing 
solution of the above linear equations. Of particular 
importance is a solution (called the {\it formal 
Baker-Akhiezer function}) of the form 
\beqn
    \Psi(\lambda) 
    \defeq \left(1 + \sum_{n=1}^\infty w_n \lambda^{-n} 
      \right) e^{\xi(\lambda)}, \nonumber
\eeqn
where the first factor is generally a formal Laurent 
series with variable coefficients $w_n = w_n(\bfx)$, 
and $\xi(\lambda)$ is given by 
\beqn
    \xi(\lambda) 
    \defeq \sum_{n=1}^\infty x_n \lambda^n.\nonumber 
\eeqn
The exponential factor is the same thing that we 
have encountered in the bosonization of free fermions. 
One can make sense of the action of PsDO's on 
$\Psi(\lambda)$ by simply extrapolating the derivation 
rule 
\beqn
    \rd_x^n e^{\xi(\lambda)} = \lambda^n e^{\xi(\lambda)}\nonumber
\eeqn
to negative powers of $\rd_x$. 

Let us now introduce the {\it Wilson-Sato operator} 
\beqn
    W \defeq 1 + \sum_{n=1}^\infty w_n \rd_x^{-n}. \nonumber
\eeqn
This enables us to express the formal Baker-Akhiezer 
function as: 
\beqn
    \Psi(\lambda) = W e^{\xi(\lambda)}. \label{eq:WPsi}
\eeqn
The linear equations for $\Psi(\lambda)$ can now 
be converted to the  equations 
\beqn
    L = W \rd_x W^{-1}, \quad 
    \frac{\rd W}{\rd x_n} = B_n W - W \rd_x^n \label{eq:L&W}
\eeqn
for $W$.  The first equation shows the relation 
between the Lax operator $L$ and the Wilson-Sato 
operator $W$ (note that $W$ is monic, hence 
invertible.) In particular, the Zakharov-Shabat 
operators can be written 
\beqn
    B_n = \left( W \rd_x^n W^{-1}\right)_{\ge 0}. \nonumber 
\eeqn
If one substitutes the Zakharov-Shabat operators 
in the second equation for $W$ by this expression, the outcome 
is the equation
\beqn
    \frac{\rd W}{\rd x_n} 
    = \left(W \rd_x^n W^{-1}\right)_{\ge 0} W 
      - W \rd_x^n 
    = - \left(W \rd_x^n W^{-1}\right)_{\le -1} W. \nonumber
\eeqn
Note that this is a {\it nonlinear} system of 
evolution equations for the coefficients $w_n$. 
This is a {\it master equation} of the KP hierarchy; 
the Lax equations can be {\it derived} from this 
nonlinear system via aforementioned relation (\ref{eq:L&W})
between $L$ and $W$.  It is this nonlinear system 
that Sato linearized on an infinite dimensional 
Grassmann manifold (Sato's {\it universal Grassmannian}) 
\cite{bib:SS,bib:SW}. 

The $\ell$-reduced hierarchy (the $\ell$-KdV hierarchy) 
is defined by the constraint 
\beqn
    (L^\ell)_{\le -1} = 0. \label{eq:redL}
\eeqn
The constraint means that $L^\ell$ be a {\it differential} 
operator, i.e., 
\beqn
    P \defeq L^\ell 
    = \rd_x^\ell + u_2 \rd_x^{\ell-2} + \ldots + u_\ell. \nonumber
\eeqn
This constraint is preserved under time evolutions, 
because the associated Lax equations 
\beqn
    \frac{\rd P}{\rd x_n} = \left[ (P^{n/\ell})_{\ge0}, P\right], \nonumber
\eeqn
for $P$ become a closed system of unconstrained evolution 
equations for the coefficients $u_2,\ldots,u_\ell$ of $P$.  
The time evolutions in $x_\ell, x_{2\ell}, \ldots$ are 
trivial, 
\beqn
    \frac{\rd P}{\rd x_{n\ell}} = [P^n, P] = 0, \nonumber
\eeqn
and, in turn, give rise to an {\it eigenvalue problem}, 
\beqn
    P \Psi(\lambda) = \lambda^\ell \Psi(\lambda).\nonumber 
\eeqn
The other time evolutions can be interpreted as 
{\it isospectral deformations} of this eigenvalue 
problem.  In the case where $\ell = 2$, this is 
exactly the well known characterization of 
the KdV hierarchy.

\subsection{Heuristic Introduction of Bogoyavlensky hierarchy} 

We here present a heuristic consideration that leads 
to a hierarchy of higher evolution equations for the 
Bogoyavlensky equation. 

The point of departure is the Lax representation 
\beqn
    \frac{\rd P}{\rd z} 
    = [P \rd_y + C, P] 
    = P \frac{\rd P}{\rd y} + [C,P] \nonumber
\eeqn
mentioned in Introduction (\ref{eq:Lax}).  Here $P$ is the Lax 
operator $\rd_x^2 + p$ of the KdV equation and $C$ is 
a first order differential operator.  This Lax equation 
is associated with the linear equations 
\beqn
    P \Psi(\lambda) = \lambda^2 \Psi(\lambda), \quad 
    \frac{\rd \Psi(\lambda)}{\rd z} = (P \rd_y + C) \Psi(\lambda).\nonumber 
\eeqn
Note that the second linear equation can be rewritten 
\beqn
    \frac{\rd \Psi(\lambda)}{\rd z} 
    = (\lambda^2 \rd_y + A) \Psi(\lambda), \nonumber
\eeqn
where 
\beqn
    A \defeq C - \frac{\rd P}{\rd y}.\nonumber 
\eeqn
Furthermore, if we assume the existence of the Wilson-Sato 
operator $W$ with which $\Psi(\lambda)$ takes the form 
\beqn
    \Psi(\lambda) = W e^{\xi(\lambda)}, \quad 
    \xi(\lambda) = x\lambda + \cdots,\nonumber 
\eeqn
the time evolution of $W$ is determined by the 
equation 
\beqn
    \frac{\rd W}{\rd z} 
    &=& P \frac{\rd W}{\rd y} + CW   \nonumber \\
    &=& P [\rd_y, W] + CW   \nonumber \\
    &=& (P \rd_y + C) W - W \rd_x^2 \rd_y.\nonumber 
\eeqn
We have used the identity $PW = W\rd_x^2$ in the 
last two lines. 

If one compares the last equation with the equations 
in the KP hierarchy, one will soon notice that the 
role of $\rd_x^n$ is now played by $\rd_x^2 \rd_y$. 
The (two-dimensional!) differential operator 
$P\rd_y + C$ should be a counterpart of $B_n$'s.  
A natural generalization of this observation is 
to consider the higher order operators $\rd_x^{2n}\rd_y$, 
$n = 1,2,\ldots$.  The associated evolution equations, 
with ``time variables'' $y_2(=z),y_4,\ldots$, will 
accordingly take such a form as 
\beqn
    \frac{\rd W}{\rd y_{2n}} 
    = \X_{2n} W - W \rd_x^{2n} \rd_y 
    \nonumber
\eeqn
where $\X_{2n}$ is a {\it two-dimensional} 
differential operator in $x$ and $y$.  
In fact, one can specify $\X_{2n}$ in more detail: 
This equation can be rewritten 
\beqn
    \frac{\rd W}{\rd y_{2n}} 
    = (\X_{2n} - P^n \rd_y) W 
    + P^n \frac{\rd W}{\rd y}, 
    \nonumber
\eeqn
and this implies that $\X_{2n} - P^n \rd_y$ 
is a differential operator in $x$ only, 
which we call $C_{2n}$: 
\beqn
    \X_{2n} = P^n \rd_y + C_{2n}. \nonumber
\eeqn

We are thus led to the evolution equations 
\beqn
    \frac{\rd W}{\rd y_{2n}} 
    &=& (P^n \rd_y + C_{2n}) W - W \rd_x^{2n}\rd_y 
    \nonumber \\
    &=& P^n \frac{\rd W}{\rd y} + C_{2n} W. \nonumber
\eeqn
Just like the Zakharov-Shabat operators $B_n$ 
of the KP hierarchy, the differential operator 
$C_{2n}$ is uniquely determined by the evolution 
equation itself: 
\beqn
    C_{2n} 
    = \left( \frac{\rd W}{\rd y_{2n}} W^{-1} 
      - P^n \frac{\rd W}{\rd y} W^{-1} \right)_{\ge 0} 
    = - \left(P^n \frac{\rd W}{\rd y} W^{-1} \right)_{\ge 0}.\nonumber 
\eeqn
Therefore the evolution equation of $W$ can be 
rewritten 
\beqn
    \frac{\rd W}{\rd y_{2n}} 
    = \left(P^n \frac{\rd W}{\rd y} W^{-1}
      \right)_{\le -1} W. \nonumber
\eeqn
These equations, along with the aforementioned 
evolution equations in $x_{2n+1}$'s, give a nonlinear 
system of evolution equations for the coefficients 
$w_n$ of $W$.  We call this system the {\it 
Bogoyavlensky hierarchy\/} or, more precisely, 
the {\it Bogoyavlensky-KdV hierarchy\/}. 

The spatial variable $y$ may be identified with 
the {\it zero-th} time variable $y_0$, because 
the corresponding operator $C_0$ is equal to 
zero so that the evolution equation for $W$ 
reduces to 
\beqn
    \frac{\rd W}{\rd y_0} = \frac{\rd W}{\rd y}.\nonumber 
\eeqn

It is straightforward to write down the associated 
linear equations.  They take two equivalent (but 
apparently different) forms.  One expression is a 
direct consequence of the evolution equation of $W$: 
\beqn
    \frac{\rd \Psi(\lambda)}{\rd y_{2n}} 
    = (P^n \rd_y + C_{2n}) \Psi(\lambda). \nonumber 
\eeqn
Another expression is the following: 
\beqn
    \frac{\rd \Psi(\lambda)}{\rd y_{2n}} 
    = (\lambda^{2n} \rd_y + \A_{2n}) \Psi(\lambda),\nonumber 
\eeqn
where 
\beqn
    \A_{2n} \defeq C_{2n} - \frac{\rd P^n}{\rd y}. \nonumber
\eeqn

\subsection{Lax equations of Bogoyavlensky hierarchy} 

Let us derive Lax equations of the Bogoyavlensky 
hierarchy.  To this end, we introduce the PsDO 
\beqn
    Q \defeq \frac{\rd W}{\rd y} W^{-1} \nonumber
\eeqn
as the {\it second} Lax operator.  This operator, 
just like $P = W \circ \rd_x^2 \circ W^{-1}$, is 
related to conjugation by the Wilson-Sato operator: 
\beqn
    \rd_y - Q = W \circ \rd_y \circ W^{-1}. \nonumber
\eeqn
The commutation relation $[\rd_y, \rd_x^2] = 0$ 
thereby implies the commutation relation 
\beqn
    [Q - \rd_y, P] = 0 \nonumber
\eeqn
or, equivalently, 
\beqn
    \frac{\rd P}{\rd y} = [Q,P]. \nonumber
\eeqn
Note that this relation itself takes the form 
of a Lax equation.

These operators have already appeared in the 
aforementioned differential operators $C_{2n}$ 
and $\A_{2n}$.  $C_{2n}$ can be indeed written 
\beqn
    C_{2n} = - (P^n Q)_{\ge 0}.\nonumber 
\eeqn
Moreover, using the Lax-type equation 
\beqn
    \frac{\rd P^n}{\rd y} = [Q, P^n] \nonumber
\eeqn
(which is a consequence of the commutation 
relations of $P$ and $Q$), one can confirm 
that $\A_{2n}$ can be written 
\beqn
    \A_{2n} = - (QP^n)_{\ge 0}. \nonumber
\eeqn

Having these operators, we can now write down 
Lax equations of the Bogoyavlensky hierarchy: 

\begin{proposition}
The operators $P$ and $Q$ satisfy the equations 
\beqn
    [\rd_{x_{2n+1}} - B_{2n+1}, P] = 0, && 
    [\rd_{x_{2n+1}} - B_{2n+1}, Q - \rd_y] = 0, 
    \nonumber \\{}
    [\rd_{y_{2n}} - P^n \rd_y - C_{2n}, P] = 0, &&
    [\rd_{y_{2n}} - P^n \rd_y - C_{2n}, Q - \rd_y] = 0, \nonumber
\eeqn
or, equivalently, 
\beqn
    \frac{\rd P}{\rd x_{2n+1}} &=& [B_{2n+1}, P], 
    \nonumber \\
    \frac{\rd Q}{\rd x_{2n+1}} 
    &=& \frac{\rd B_{2n+1}}{\rd y} + [B_{2n+1}, Q],  
    \nonumber \\
    \frac{\rd P}{\rd y_{2n}}
    &=& P^n \frac{\rd P}{\rd y} + [C_{2n}, P], 
    \nonumber \\
    \frac{\rd Q}{\rd y_{2n}} 
    &=& P^n \frac{\rd Q}{\rd y} + \frac{\rd C_{2n}}{\rd y} 
      + [C_{2n}, Q]. \nonumber
\eeqn
\end{proposition}

\proof
Let us prove the equations for the $y_{2n}$-derivatives; 
the proof of the other equations are parallel and even 
simpler.  First, differentiating $P = W \rd_x^2 W^{-1}$ 
gives 
\beqn
    \frac{\rd P}{\rd y_{2n}} 
    &=& \frac{\rd W}{\rd y_{2n}} \rd_x^2 W^{-1} 
      - W \rd_x^2 W^{-1} \frac{\rd W}{\rd y_{2n}} W^{-1} 
    \nonumber \\
    &=& \left[\frac{\rd W}{\rd y_{2n}} W^{-1}, P\right]. 
    \nonumber 
\eeqn
We can now use the evolution equations 
\beqn
    \frac{\rd W}{\rd y_{2n}} 
    = P^n Q W + C_{2n} W 
    \nonumber 
\eeqn
and the commutation relation $[Q,P] = \rd P/\rd y$, 
and find that 
\beqn
    \frac{\rd P}{\rd y_{2n}} 
    &=& [P^n Q + C_{2n}, P] 
    \nonumber \\
    &=& P^n \frac{\rd Q}{\rd y} + [C_{2n}, P]. 
    \nonumber 
\eeqn
Similarly, from the identity 
$Q - \rd_y = - W \rd_y W^{-1}$, 
\beqn
    \frac{\rd Q}{\rd y_{2n}} 
    &=& \left[\frac{\rd W}{\rd y_{2n}} W^{-1}, 
        Q - \rd_y\right] 
    \nonumber \\
    &=& [P^n Q + C_{2n}, Q - \rd_y] 
    \nonumber \\
    &=& [P^n, Q] Q + \frac{\rd P^n Q}{\rd y} 
      + \frac{\rd C_{2n}}{\rd y} + [C_{2n}, Q] 
    \nonumber \\
    &=& [P^n, Q] Q + \frac{\rd P^n}{\rd y} Q 
      + P^n \frac{\rd Q}{\rd y} 
      + \frac{\rd C_{2n}}{\rd y} + [C_{2n}, Q]. 
    \nonumber 
\eeqn
The first two terms in the last line cancel each other, 
again because of the commutation relation of $Q$ and $P$.  
This completes the proof. \qed

\subsection{Formulation of $\ell$-Bogoyavlensky hierarchy}

We are now in a position to formulate the 
$\ell$-Bogoyavlensky hierarchy for a general 
value of $\ell \ge 2$. This hierarchy is an 
extension of the $\ell$-reduced KP hierarchy 
with additional independent variables $\bfy=(y,\cbfy)$,
where $y = y_0$ and 
$\cbfy = (y_\ell,y_{2\ell},\ldots)$.
The coefficients $w_n$ of the Wilson-Sato operator 
\beqn
    W \defeq 1 + \sum_{n=1}^\infty w_n \rd_x^{-n}\nonumber
\eeqn
are understood to be functions of these variables also,
$w_n=w_n(\bfx,\bfy).$ Assume that $W$ satisfy
the $\ell$-reduced constraints, cf.(\ref{eq:redL}):
\beqn
\left(W\rd_x^\ell W^{-1}\right)_{\le -1}=0,\nonumber
\eeqn
and define the differential operator
\beqn
P\defeq
W\rd_x^\ell W^{-1}.\nonumber
\eeqn

\begin{itemize}
\item 
The most fundamental part of this hierarchy  is the 
evolution equations 
\beqn
    \frac{\rd W}{\rd x_n} 
    &=& - \left(W \rd_x^n W^{-1}\right)_{\le -1} W, 
    \nonumber \\
    \frac{\rd W}{\rd y_{n\ell}} 
    &=& \left(W \rd_x^{n\ell}W^{-1}\frac{\rd W}{\rd y} 
        W^{-1}\right)_{\le -1} W. \nonumber
\eeqn
of $W$. These equations can also be written 
\beqn
    \frac{\rd W}{\rd x_n} &=& B_n W - W \rd_x^n, 
    \label{eq:Sato-x} \\
    \frac{\rd W}{\rd y_{n\ell}} 
    &=& {W \rd_x^{n\ell} W^{-1}} \frac{\rd W}{\rd y} + C_{n\ell} W
    \label{eq:Sato-y}  \\
    &=& ({W \rd_x^{n\ell} W^{-1}} \rd_y + C_{n\ell})W 
    - W \rd_x^{n\ell}\rd_y, \nonumber
\eeqn
where $B_n$ and $C_{n\ell}$ are differential operators 
defined by 
\beqn
    B_n &\defeq& \left(W \rd_x^n W^{-1} \right)_{\ge 0},\nonumber\\
    C_{n\ell} &\defeq& - \left(W \rd_x^{n\ell} W^{-1}
              \frac{\rd W}{\rd y}W^{-1}\right)_{\ge 0}.
              \label{eq:defC}
\eeqn
\item 
The evolution equations of $W$ in $y_{n\ell}$'s 
can also be rewritten 
\beqn
    \frac{\rd W}{\rd y_{n\ell}} 
    = \frac{\rd W}{\rd y} \rd_x^{n\ell} 
    + \A_{n\ell} W, \nonumber
\eeqn
where 
\beqn
    \A_{n\ell} 
    \defeq - (QP^n)_{\ge 0}
    = C_{n\ell} - \frac{\rd P^n}{\rd y}.\label{eq:defA}   
\eeqn
\item 
Lax equations of this hierarchy are described by $P$ and the 
following  PsDO: 
\beqn
    Q \defeq \frac{\rd W}{\rd y} W^{-1}.\nonumber 
\eeqn
They satisfy the commutation relation 
\beqn
    [Q,P] = \frac{\rd P}{\rd y}. \nonumber
\eeqn
$B_n$ and $C_{n\ell}$ can now be expressed as 
\beqn
    B_n = (P^{n/\ell})_{\ge 0}, \quad 
    C_{n\ell} = - (P^n Q)_{\ge 0}. \nonumber
\eeqn
$P$ and $Q$ satisfy the Lax equations: 
\beqn
    \frac{\rd P}{\rd x_n} &=& [B_n, P], 
    \nonumber \\
    \frac{\rd Q}{\rd x_n} 
    &=& \frac{\rd B_n}{\rd y} + [B_n, Q],  
    \nonumber \\
    \frac{\rd P}{\rd y_{n\ell}} 
    &=& P^n \frac{\rd P}{\rd y} + [C_{n\ell}, P], 
    \nonumber \\
    \frac{\rd Q}{\rd y_{n\ell}} 
    &=& P^n \frac{\rd Q}{\rd y} + \frac{\rd C_{n\ell}}{\rd y} 
      + [C_{n\ell}, Q]. \nonumber
\eeqn
\item 
The formal Baker-Akhiezer function 
\beqn 
    \Psi(\lambda) \defeq W e^{\xi(\lambda)}\label{eq:Psi&W}
\eeqn
satisfies the linear equations
\beqn
    P \Psi(\lambda) = \lambda^\ell \Psi(\lambda), &&
    Q \Psi(\lambda) = \frac{\rd \Psi(\lambda)}{\rd y}, 
    \nonumber \\
    \frac{\rd \Psi(\lambda)}{\rd x_n} = B_n \Psi(\lambda), && 
    \frac{\rd \Psi(\lambda)}{\rd y_{n\ell}} 
    = \left(P^n \rd_y + C_{n\ell}\right) \Psi(\lambda).\label{eq:linear} 
\eeqn
The last equation can also be written 
\beqn
    \frac{\rd \Psi(\lambda)}{\rd y_{n\ell}} 
    = \left( \lambda^{n\ell} \rd_y + \A_{n\ell}\right) \Psi(\lambda).
    \label{eq:lin_Psi_D}
\eeqn
\end{itemize}

\bigskip
\textbf{Example 6}
Let us consider the case $\ell=2$.
The Lax operator has the form
$P=\rd_x^2+u$ where $u=-2w_{1x}.$
We can write down $C_2=A_2+\frac{\rd P}{\rd y}$ as follows:
\beqn
C_2=c_1\rd_x+c_2,\quad c_1=-w_{1y},\quad
c_2=w_{1y} w_1-w_{2y}-2w_{1xy}.
\eeqn
The evolution equation with respect to $y_2$ reads
\beqn
\frac{\rd P}{\rd y_2}=P\frac{\rd P}{\rd y}+[C_2,P].\label{eq:Lax_y2}
\eeqn
One can calculate the right-hand side of (\ref{eq:Lax_y2}) as follows:
\beqn
&&P\frac{\rd P}{\rd y}+[C_2,P]\\
&=&(2u_{xy}-c_{1xx}-2c_{2x})\rd_x
+u_{xxy}+uu_y+c_1 u_x-c_{2xx},\label{eq:Ry2}
\eeqn
where the coefficients of $\rd_x^2$ 
cancel trivially because $u=-2w_{1x}$.

Here we use the following relation which 
is a direct consequence of the $2$-reduced condition
\beqn
w_{1xx} + 2w_{2x} -2w_1 w_{1x}=0.
\eeqn
Using this, we now have 
$
c_{2x}=-\frac{3}{2}w_{1xxy}.
$
This finally leads to 
\beqn
\frac{\rd u}{\rd y_2}=\frac{1}{4}u_{xxy}+vu_x+uu_y\quad
\mbox{where}\quad v\defeq -w_{1y}.
\eeqn

\textbf{Caution!}
The expression $w_{2x}$, for example, does not mean $\rd_x^2 w$ but $\rd_x w_2$. 

\bigskip
\textbf{Example 7}
We shall demonstrate the third order
flow $y_3$ in case of $\ell=3$.
The Lax operator $P=\rd_x^3+U\rd_x+V$ is given as:
\beqn
U=-3w_{1x},\quad 
V=-3w_{1xx}-3w_{2x}+3w_1 w_{1x}.\label{eq:UV}
\eeqn
Operator $C_3=c_1\rd_x^2+c_2\rd_x+c_3$ is 
given explicitly as: 
\beqn
c_1&=&-w_{1y},\quad c_2=w_{1y}w_1-w_{2y}-3w_{1xy},\label{eq:c2v}\\
c_3&=&-w_{3y}-w_{1y}w_1^2+w_1 (w_{2y}+3w_{1xy})+w_{1y}(w_2
+5w_{1x})-3w_{1xxy}-3w_{2xy}.\label{eq:c3}
\eeqn
Now we recall the $3$-reduced condition.
In particular, 
the coefficient of $\rd_x^{-1}$ in $W\rd_x^3 W^{-1}$
should vanish identically. This condition is written as:  
\beqn
w_{3x}=-\textstyle\frac{1}{3}w_{1xxx}-w_{2xx}+w_{1x}^2+w_{1x}w_2
+w_1 w_{1xx}+w_1 w_{2x}-w_{1x}w_1^2.\label{eq:3red}
\eeqn

Using (\ref{eq:UV}),(\ref{eq:c2v}), and (\ref{eq:c3}),
it is simple to see that the right-hand side of 
evolution equation, 
\beqn
\rd_{y_3}P=P\frac{\rd P}{\rd y}+[C_3,P], \label{eq:y_3}
\eeqn 
is of order at most two. 
To verify that the right-hand side
is actually of first order, we eliminate $w_3$
by using equation (\ref{eq:3red}).
Thus, it should be emphasized that the $3$-reduced condition
is crucial to introduce flow $y_3$ by (\ref{eq:y_3}). 

After eliminating $w_3$,
we have the following explicit forms of evolution equations:
\beqn
\rd_{y_3}U
&=&6w_{1y}w_{2xx}-w_{1xxxxy}+9w_{1xy}w_{2x}
-2w_{2xxxy}-12 w_{1y}w_{1x}^2-15w_{1xy}w_1w_{1x}\nonumber\\
&&+6w_{1x}w_{2xy}-9w_{1y}w_1w_{1xx}+12w_{1xy}w_{1xx}
+9w_{1xxy}w_{1x}+2w_1w_{1xxxy}\nonumber\\
&&+5w_{1y}w_{1xxx}+3w_{1xx}w_{2y},\nonumber\\
\rd_{y_3}V
&=&-9w_{1xy}w_{1x}^2-18w_{1xy}w_1w_{1xx}+3w_{1xxx}w_{2y}
+3w_{1x}w_{2xxy}-3w_{1x}^2w_{2y}
\nonumber\\
&&-15w_{1y}w_{1x}w_{1xx}
+9w_{1xy}w_{2xx}
+9w_{1xxy}w_{2x}
-12w_{1xxy}w_1w_{1x}
\nonumber\\
&&+3w_{1y}w_{2xxx}
-6w_{1y}w_1w_{1xxx}
+2w_{1y}w_{1xxxx}
+7w_{1xy}w_{1xxx}
+9w_{1xxy}w_{1xx}\nonumber\\
&&
+3w_{1xxxy}w_{1x}
+9w_{1xx}w_{2xy}
-3w_{2y}w_{1}w_{1xx}
+12w_{1y}w_{1}w_{1x}^2
+3w_{1y}w_{1}^2w_{1xx}\nonumber\\
&&
-9w_{2x}w_{1}w_{1xy}
+9w_{2x}w_{2xy}
+3w_{2y}w_{2xx}
-9w_{2x}w_{1y}w_{1x}
-3w_{1y}w_{1}w_{2xx}\nonumber\\
&&-9w_{1}w_{1x}w_{2xy}
+9w_{1}^2w_{1x}w_{1xy}
-w_{2xxxxy}
-\textstyle\frac{1}{3}w_{1xxxxxy}
+w_1 w_{1xxxxy}.\nonumber
\eeqn 

\subsection{Explicit forms of operators $\A_n$}\label{subsec:An}
We give a supplementary discussion on the forms of 
differential operators $A_n$.
The differential operators $\A_{n\ell}$ and $C_{n\ell}$ are
written in terms of $W$ (\ref{eq:defA}),(\ref{eq:defC}) 
and hence 
in the coefficients $w_i$'s of $W.$
In particular, we have 
\beqn
\A_n=\sum_{i=1}^{n} a_{i}^{(n)}\rd_x^{n-i}
=-\left(\frac{\rd W}{\rd y}\rd_x^n W^{-1}\right)_{\ge 0}.
\nonumber
\eeqn
Here we shall recursively determine 
the coefficients $a_i^{(n)}.$
If we cast $\Psi(\lambda)=We^{\xi(\lambda)}$ 
into linear equation (\ref{eq:lin_Psi_D}), then we obtain
\beqn
\frac{\rd W}{\rd y_n}
=\sum_{i=1}^{n} a_{i}^{(n)}(\rd_x+\lambda)^{n-i}W
+\lambda^n\frac{\rd W}{\rd y}.\nonumber
\eeqn

Taking the polynomial part of the equation above with respect to $\lambda,$
we obtain 
\beqn
\sum_{i=1}^{n} a_{i}^{(n)}(\rd_x+\lambda)^{n-i}\left(\sum_{j=0}^n w_j\lambda^{-j}\right)
+\sum_{j=1}^n(\rd_y w_j)\lambda^{n-j}=0,
\quad w_0\defeq 1.\nonumber
\eeqn
Equating coefficients of $\lambda^{n-1}$ in the equation above,
we obtain $a^{(n)}_1=-\rd_y w_1.$
Moreover, if $a_1^{(1)},\ldots,a_{i-1}^{(n)}$ are already known, then
$a_{i}^{(n)}$ can be uniquely determined
by the coefficients of $\lambda^{n-i}$.
In particular, we have $a_2^{(n)}=w_1\rd_y w_1-\rd_y w_2.$
Note that $a_i^{(n)}$ is a differential polynomial 
of $w_1,\ldots,w_i$ with respect to  $\rd_x$ and $\rd_y.$

\bigskip
\textbf{Example 8}
For any $n$ we have 
$a_1^{(n)}=-w_{1y},\;a_2^{(n)}=w_{1y}w_1-w_{2y}.$
We shall give the formulae for $n\leq 5$:
\beqn
a_3^{(3)}&=&-w_{3y}-w_{1y}w_1^2+w_1 w_{2y}+w_{1y}w_2+2w_{1y}w_{1x},
\nonumber\\
a_3^{(4)}&=&w_1w_{2y}-w_{1y}w_1^2+3w_{1y}w_{1x}-w_{3y}+w_{1y}w_2,
\nonumber\\
a_4^{(4)}&=&-w_{4y}-w_1^2 w_{2y}+w_{1y} w_1^3-5 w_{1x}w_{1y}w_1
+w_1w_{3y}-2w_2w_{1y}w_1\nonumber\\
&&+w_2w_{2y}+2w_{1x}w_{2y}+w_{1y}w_3
+3w_{1y}w_{2x}+3w_{1y}w_{1xx},\nonumber\\
a_3^{(5)}&=&-w_{1y}w_1^2+w_1 w_{2y}-w_{3y}+4w_{1y}w_{1x}
+w_{1y}w_2,\nonumber\\
a_4^{(5)}&=&6w_{1y}w_{1xx}+w_{1y}w_1^3-w_1^2w_{2y}
+w_1w_{3y}-7w_{1x}w_{1y}w_1-2w_2w_{1y}w_1\\
&&+3w_{1x}w_{2y}+w_2w_{2y}-w_{4y}+4w_{1y}w_{2x}+w_{1y}w_3,\nonumber\\
a_5^{(5)}&=&-w_{5y}+w_{1y}w_4+w_3w_{2y}+3w_{1xx}w_{2y}+3w_{2x}w_{2y}
+4w_{1y}w_{1xxx}+3w_2w_{1y}w_1^2\nonumber\\
&&-2w_2w_1w_{2y}-6w_2w_{1y}w_{1x}+9w_{1x}w_{1y}w_1^2
-5w_{1x}w_1w_{2y}-2w_3w_{1y}w_1-9w_{1xx}w_{1y}w_1\nonumber\\
&&-7w_{2x}w_{1y}w_1+6w_{1y}w_{2xx}+4w_{1y}w_{3x}+w_2w_{3y}
-w_{1y}w_2^2+2w_{1x}w_{3y}-8w_{1y}w_{1x}^2\nonumber\\
&&-w_{1y}w_1^4+w_1^3w_{2y}-w_1^2w_{3y}+w_1w_{4y}.
\eeqn

\section{Reproducing bilinear equations in Lax formalism}
\label{section:LaxToBilin}
\setcounter{equation}{0}

\subsection{Dual formal Baker-Akhiezer function}

We define the {\it dual} $\Psi^*(\lambda)$ of the formal 
Baker-Akhiezer function $\Psi(\lambda)$ as follows: 
\beqn
    \Psi^*(\lambda) \defeq W^{*-1} e^{-\xi(\lambda)}.\nonumber 
\eeqn

\begin{proposition}
$\Psi^*(\lambda)$ satisfies the linear equations 
\beqn
    P^* \Psi(\lambda) = \lambda^\ell \Psi^*(\lambda), &&
    Q^* \Psi(\lambda) = - \frac{\rd \Psi^*(\lambda)}{\rd y}, 
    \nonumber \\
    \frac{\rd \Psi^*(\lambda)}{\rd x_n} 
    = - B_n^* \Psi^*(\lambda), &&
    \frac{\rd \Psi^*(\lambda)}{\rd y_{n\ell}}
    = \left(P^{*n} \rd_y - \A_{n\ell}\right) \Psi^*(\lambda). \nonumber
\eeqn
The last equation can also be written 
\beqn
    \frac{\rd \Psi^*(\lambda)}{\rd y_{n\ell}} 
    = \left( \lambda^{n\ell} \rd_y - C_{n\ell}^*
      \right) \Psi^*(\lambda). \nonumber
\eeqn
\end{proposition}

\proof
The first and second equations are immediate from 
the operator relations 
\beqn
    P^* W^{*-1} &=& - W^{*-1} \rd_x, 
    \nonumber \\ 
    Q^* W^{*-1} 
    &=& W^{*-1} \frac{\rd W}{\rd y} W^{*-1} 
    = - \frac{\rd W^{*-1}}{\rd y}. 
    \nonumber 
\eeqn
The third equation can be derived as follows: 
\beqn
    \frac{\rd W^{*-1}}{\rd x_n}
    &=& - W^{*-1} \frac{\rd W^*}{\rd x_n} W^{*-1} 
    \nonumber \\
    &=& - W^{*-1} \left(B_n W - W \rd_x^n\right)^* W^{*-1} 
    \nonumber \\
    &=& - B_n^* W^{*-1} + W^{*-1} (-\rd_x)^n. 
    \nonumber 
\eeqn
Similarly, 
\beqn
    \frac{\rd W^{*-1}}{\rd y_{n\ell}} 
    &=& - W^{*-1} \frac{\rd W^*}{\rd y_{n\ell}} W^{*-1} 
    \nonumber \\
    &=& - W^{*-1} \left(P^n \frac{\rd W}{\rd y_{n\ell}} 
          + C_{n\ell} W\right)^* W^{*-1} 
    \nonumber \\
    &=& \frac{\rd W^{-1}}{\rd y} (-\rd_x)^{n\ell} 
      - C_{n\ell} W^{*-1}, 
    \nonumber 
\eeqn
which implies the fourth equation.  The last equation 
can be readily derived from the fourth equation. 
\qed

\subsection{Residue formula of product of PsDO's}

The following lemma \cite{bib:Kas,bib:DJKM} is a clue 
for deriving the bilinear equation that $\Psi(\lambda)$ 
and $\Psi^*(\lambda)$ satisfy. 

\begin{lemma}\label{lemma:integral}
For any pair $A,B$ of PsDO's and non-negative integer $j$, 
\beqn
    \frac{1}{2\pi i}\oint_{\lambda=\infty} 
    \left( \rd_x^j A e^{x\lambda} \right) 
    \left( B e^{-x\lambda} \right) d\lambda 
    = (-1)^{j+1} (BA^*)_{-j-1}. \nonumber
\eeqn 
In particular, if $A$ and $B$ are PsDO's of zero-th 
order of the form $A = a_0 + O(\rd_x^{-1})$ and 
$B = b_0 + O(\rd_x^{-1})$, 
\beqn
    BA^* = b_0 a_0 \quad \Longleftrightarrow \quad 
    (\forall j \ge 0) \ 
    \oint_{\lambda=\infty} 
    \left( \rd_x^j A e^{x\lambda} \right) 
    \left( B e^{-x\lambda} \right) d\lambda 
    = 0.\nonumber     
\eeqn
\end{lemma}

\proof 
Let $A$ and $B$ be written $A = \sum_n a_n \rd_x^n$ 
and $B = \sum_n b_n \rd_x^n$.  Since 
\beqn
    \rd_x^j \circ A 
    = \sum_m \sum_{k \ge 0} \binom{j}{k} 
      a_m^{(k)} \rd_x^{j+m-k}, 
    \nonumber 
\eeqn
the contour integral can be calculated as: 
\beqn
    && \frac{1}{2\pi i}\oint_{\lambda=\infty} 
      \left(\rd_x^k A e^{x\lambda}\right) 
      \left(B e^{-x\lambda}\right) d\lambda  
    \nonumber \\
    &=& \frac{1}{2\pi i} \oint_{\lambda=\infty} 
      \sum_{j,k} \binom{j}{k} a_m^{(k)} \lambda^{j+m-k} 
      \cdot \sum_n b_n (-\lambda)^n d\lambda 
    \nonumber \\
    &=& \sum_{k-m-n-1=j} \binom{j}{k} (-1)^n a_m^{(k)} b_n. 
    \nonumber 
\eeqn
On the other hand, 
\beqn
    BA^* &=& \sum_{m,n} b_n \rd_x^n (-\rd_x)^m \circ a_m 
    \nonumber \\
    &=& \sum_{m,n,k} \binom{m+n}{k} 
      (-1)^m a_m^{(k)} \rd_x^{m+n-k} 
    \nonumber \\
    &=& \sum_{m,n,k} \binom{k-m-n-1}{k} 
      (-1)^{k-m} a_m^{(k)} b_n \rd_x^{m+n-k} 
    \nonumber \\
    &=& \sum_j \sum_{k-m-n-1=j} \binom{j}{k} 
      (-1)^n a_m^{(k)} b_n (-\rd_x)^{-j-1}. 
    \nonumber 
\eeqn
Therefore the coefficient of $\rd_x^{-j-1}$  coincides 
with the countour integral.  The lemma is thus proven. 
\qed 

\subsection{Bilinearization of linear equations}

Let us slightly change the notation, namely, we make explicit
the functional dependence on $\bfx$ 
and $\bfy,$ 
e.g., $\Psi(\bfx,\bfy,\lambda)$, 
$\Psi^*(\bfx,\bfy,\lambda)$ and $\xi(\bfx,\lambda)$ 
stand for $\Psi(\lambda)$, $\Psi^*(\lambda)$ and 
$\xi(\lambda)$ in the previous notation.  

We now apply Lemma \ref{lemma:integral} to the case 
where $A = W$ and $B = W^{*-1}$. This eventually leads 
to a bilinear equation for $\Psi(\bfx,\bfy,\lambda)$ 
and $\Psi^*(\bfx,\bfy,\lambda)$ as follows: 
\begin{enumerate}
\item By the lemma applied to $A = W$  and $B = W^{*-1}$ 
gives the bilinear  identities 
\beqn
    \oint_{\lambda=\infty} 
    \Bigl(\rd_x^j \Psi(\bfx,\bfy,\lambda)\Bigr) 
    \Psi^*(\bfx,\bfy,\lambda) d\lambda 
    = 0 \nonumber
\eeqn
for $j \ge 0$ (note that $x_2 \lambda^2 + x_3\lambda^3 + \cdots$ 
in $e^{\pm \xi(\bfx,\lambda)}$ cancel out).  Thus, 
we have 
\beqn
    \oint_{\lambda=\infty} 
    \Bigl(R \Psi(\bfx,\bfy,\lambda)\Bigr) 
    \Psi^*(\bfx,\bfy,\lambda) d\lambda 
    = 0 \nonumber
\eeqn
for any differential operator 
$R = \sum_{n \ge 0} r_n(\bfx,\bfy) \rd_x^n$ in $x.$
\item 
Iteration of the evolution equation of 
$\Psi(\bfx,\bfy,\lambda)$ gives rise to 
higher order equations of the form 
\beqn 
    \rd_{x_1}^{\alpha_1} \rd_{x_2}^{\alpha_2} 
    \cdots \Psi(\bfx,\bfy,\lambda) 
    = B_{\alpha_1,\alpha_2,\ldots} \Psi(\bfx,\bfy,\lambda)\nonumber 
\eeqn
for $\alpha_1,\alpha_2,\ldots \ge 0$, 
$B_{\alpha_1,\alpha_2,\ldots}$ being 
a differential operator in $x$.  
\item 
Combining these equations with the last bilinear identity, 
we obtain the bilinear equations 
\beqn
    \oint_{\lambda=\infty} 
    \Bigl( \rd_{x_1}^{\alpha_1} \rd_{x_2}^{\alpha_2} 
      \cdots \Psi(\bfx,\bfy,\lambda)\Bigr) 
    \Psi^*(\bfx,\bfy,\lambda) d\lambda 
    = 0 \nonumber
\eeqn
for $\alpha_1,\alpha_2,\ldots \ge 0$. 
\item 
These bilinear equations can be cast into a generating 
function: Introduce an infinite number of new variables 
$\bfa = (a_1,a_2,\ldots)$ , and sum up the bilinear 
equations over $\alpha_1,\alpha_2,\ldots \ge 0$ 
with the weight $a_1^{\alpha_1} a_2^{\alpha_2} 
\cdots / \alpha_1! \alpha_2! \cdots$.  The outcome is 
a single bilinear equation of the form 
\beqn
    \oint_{\lambda=\infty} \Psi(\bfx + \bfa,\bfy,\lambda) 
    \Psi^*(\bfx,\bfy,\lambda) d\lambda 
    = 0.\nonumber 
\eeqn
\item 
Finally, rewriting $\bfx + \bfa$ to $\bfx'$, we obtain 
the bilinear identity 
\beqn
    \oint_{\lambda=\infty} \Psi(\bfx',\bfy,\lambda) 
    \Psi^*(\bfx,\bfy,\lambda) d\lambda 
    = 0. \nonumber
\eeqn
\end{enumerate}

\remark 
Note that the status of $x_\ell,x_{2\ell},\ldots$ 
are somewhat different from others:  Since $w_n$ 
do not depend on these variables, they can appear 
in $e^{\pm\xi(\bfx,\lambda)}$ only.  The net effect of 
these variables is thereby insertion of the exponential 
factor 
\beqn
    \exp\Bigl(\sum_{n=1}^\infty 
    (x_{n\ell}' - x_{n\ell})\lambda^{n\ell}\Bigr) \nonumber     
\eeqn
in the contour integral of the bilinear equations. 
Therefore the last bilinear equation can be further 
reduced to the bilinear equations 
\beqn
    \oint_{\lambda=\infty} \lambda^{j\ell} 
    \Psi(\bfx',\bfy,\lambda) \Psi^*(\bfx,\bfy,\lambda)
    |_{x'_\ell=x'_{2\ell}=\cdots=x_\ell=x_{2\ell}=\cdots=0} d\lambda 
    = 0 \nonumber
\eeqn
for $j \ge 0$. 
\bigskip

This is not the end of the story.  What we have 
done is simply to bilinearize the KP-like part of 
the $\ell$-Bogoyavlensky hierarchy.  The other 
evolution equations  $y_{n\ell}$'s still remain 
to be bilinearized. 

Bilinearization of the evolution equations in 
$y_{n\ell}$'s can be achieved by the following steps, 
which are almost parallel to the previous calculations: 
\begin{enumerate}
\item 
We rewrite the evolution equations as 
\beqn
    (\rd_{y_{n\ell}} - \lambda^{n\ell}\rd_{y }) 
    \Psi(\bfx,\bfy,\lambda) 
    = \A_{n\ell} \Psi(\bfx,\bfy,\lambda)\nonumber 
\eeqn
and again do iteration.  This yields the higher order 
equations 
\beqn
 && (\rd_{y_\ell} - \lambda^\ell \rd_{y })^{\beta_1} 
    (\rd_{y_{2\ell}} - \lambda^{2\ell}\rd_{y })^{\beta_2}
    \cdots \Psi(\bfx,\bfy,\lambda) 
 \nonumber \\
 &=& \A_{\beta_1,\beta_2,\ldots} \Psi(\bfx,\bfy,\lambda) \nonumber
\eeqn
for $\beta_1,\beta_2,\ldots \ge 0$, $\A_{\beta_1,\beta_2,\ldots}$ 
being a differential operator in $x$.  
\item 
Combining these equations with the bilinear equations that 
we have derived above, we now obtain the bilinear equations 
\beqn
    \oint_{\lambda=\infty} && 
    \Bigl( (\rd_{y_\ell} - \lambda^\ell \rd_{y })^{\beta_1} 
    (\rd_{y_{2\ell}} - \lambda^{2\ell}\rd_{y })^{\beta_2} 
    \cdots \Psi(\bfx',\bfy,\lambda) \Bigr) 
    \nonumber \\
    &\times& \Psi^*(\bfx,\bfy,\lambda) d\lambda 
    = 0. \nonumber
\eeqn
\item
Introduce new variables $\bfb = (b_\ell,b_{2\ell},\ldots)$ 
and sum up the bilinear identities over 
$\beta_1,\beta_2,\ldots \ge 0$ with the weight 
$b_\ell^{\beta_1} b_{2\ell}^{\beta_2} \cdots 
/\beta_1!\beta_2! \cdots$.  This sum contains power series 
of $\rd_{y_{n\ell}} - \lambda^{n\ell} \rd_{y }$, 
which is essentially an exponential:
\beqn
    \sum_{\beta \ge 0} \frac{b^\beta}{\beta!} 
      \prod_{n=1}^\infty
      (\rd_{y_{n\ell}} - \lambda^{n\ell}\rd_{y })^{\beta_n}
    = \exp \Bigl(\sum_{n=1}^\infty b_{n\ell}
    (\rd_{y_{n\ell}} - \lambda^{n\ell}\rd_{y})\Bigr). \nonumber
\eeqn 
The action of this exponential operator can be easily 
understood by the obvious identity 
\beqn
    \exp\Bigl( \sum_{n= 1}^\infty b_{n\ell} 
    (\rd_{y_{n\ell}} - \lambda^{n\ell}\rd_{y })
    \Bigr) f(y ,\cbfy) 
    = f\Bigl(y  - \sum_{n=1}^\infty b_{n\ell}\lambda^{n\ell}, 
      \cbfy + \bfb\Bigr) \nonumber
\eeqn
that holds true for any function $f(y,\cbfy)$.  
We thus obtain the bilinear identity 
\beqn
    \oint_{\lambda=\infty} 
    \Psi\Bigl(\bfx',y- \eta(\bfb,\lambda), 
    \cbfy + \bfb,\lambda\Bigr) 
    \Psi^*(\bfx,\bfy,\lambda) d\lambda
    = 0,\nonumber
\eeqn
where for indeterminates $\bfb=(b_\ell,b_{2\ell},\ldots)$ we set
\beqn
\eta(\bfb,\lambda)\defeq
\sum_{n=1}^\infty b_{n\ell}\lambda^{n\ell}.\label{eq:def_eta}
\eeqn
\item 
These calculations can be repeated, now starting with 
the linear equations for $\Psi^*(\bfx,\bfy,\lambda)$. 
This yields a bilinear equation with the variables 
$\bfy=(y,\cbfy)$ in $\Psi^*$ being shifted as
\beqn
    (y,\cbfy) \quad \longrightarrow \quad 
    \Bigl(y - \eta(\bfc,\lambda), 
      \cbfy + \bfc\Bigr) \nonumber
\eeqn
where $\bfc = (c_\ell,c_{2\ell},\ldots)$ are newly 
introduced variables. 
\end{enumerate}
We thus eventually find the following result: 

\begin{theorem}
The formal Baker-Akhiezer functions $\Psi(\bfx,\bfy,\lambda)$ 
and $\Psi^*(\bfx,\bfy,\lambda)$ satisfy the bilinear 
equation 
\beqn
    \oint_{\lambda=\infty} && 
    \Psi\Bigl(\bfx', y- \eta(\bfb,\lambda), 
      \cbfy + \bfb,\lambda\Bigr) 
    \nonumber \\
    &\times& 
    \Psi^*\Bigl(\bfx, y - \eta(\bfc,\lambda), 
      \cbfy + \bfc,\lambda\Bigr) d\lambda
    = 0. \label{eq:Psi_bil}
\eeqn
Here $\bfx$, $\bfx'$, $y $, $\bfb$ and $\bfc$ are 
understood to be independent variables.  
\end{theorem}


\remark
\begin{enumerate}
\item Conversely, one can reproduce the evolution 
equations of $W$ etc. from these bilinear equations.  
\item The preceding remark on the status of 
$x_\ell,x_{2\ell},\ldots$ also applies to this 
bilinear equation.  Thus the bilinear equation 
holds true if any nonnegative power of $\lambda^\ell$ 
is inserted.  Accordingly, by taking a linear 
combination of those equations, one eventually 
obtains the slightly generalized (but actually 
equivalent) form 
\beqn
    \oint_{\lambda=\infty} 
    && f(\lambda^\ell) 
    \Psi\Bigl(\bfx', y- \eta(\bfb,\lambda), 
      \cbfy + \bfb,\lambda\Bigr) 
    \nonumber \\
    &\times& 
    \Psi^*\Bigl(\bfx, y - \eta(\bfc,\lambda), 
      \cbfy + \bfc,\lambda\Bigr) d\lambda
    = 0. \nonumber
\eeqn
of the bilinear equation that holds for any power 
series $f(\lambda^\ell) = \sum_{n\ge 0} f_n \lambda^{n\ell}$. 
\end{enumerate}

\subsection{$\tau$ function and Hirota bilinear equations}

We can define the $\tau$ function in the same way as 
the case of the KP hierarchy. Namely, the $\tau$ function 
is a function $\tau = \tau(\bfx,\bfy)$ that 
satisfies the equations 
\beqn
    \Psi(\bfx,\bfy,\lambda) 
    &=& \frac{\tau(\bfx - \epsilon(\lambda),\bfy)}
        {\tau(\bfx,\bfy)} e^{\xi(\bfx,\lambda)}, 
    \label{eq:Psi&tau}\\
    \Psi^*(\bfx,\bfy,\lambda) 
    &=& \frac{\tau(\bfx + \epsilon(\lambda),\bfy)}
        {\tau(\bfx,\bfy)} e^{-\xi(\bfx,\lambda)},\nonumber
\eeqn
where 
\beqn
    \epsilon(\lambda) 
    \defeq \left(\frac{1}{\lambda},\frac{1}{2\lambda^2},\ldots,
      \frac{1}{n\lambda^n},\ldots \right).\nonumber 
\eeqn
Note that this definition allows the indeterminacy 
\beqn
    \tau(\bfx,\bfy) \ \longrightarrow \ 
    f(\bfy) \tau(\bfx,\bfy) \nonumber
\eeqn
with $f(\bfy)$ being an arbitrary function of 
$y$ and $\cbfy$ only.
Equation (\ref{eq:Psi_bil}) now turns into 
the equation 
\beqn
    \oint_{\lambda=\infty} && 
       \exp\Bigl(\xi(\bfx'-\bfx,\lambda)\Bigr) 
    \nonumber \\
    &\times&
       \frac{\tau\Bigl( \bfx' - \epsilon(\lambda), 
       y - \eta(\bfb,\lambda), \cbfy + \bfb\Bigr)} 
       {\tau\Bigl(\bfx', y - \eta(\bfb,\lambda), 
       \cbfy + \bfb\Bigr)}
       \frac{\tau\Bigl(\bfx + \epsilon(\lambda), 
       y - \eta(\bfc,\lambda), \cbfy + \bfc\Bigr)} 
       {\tau\Bigl(\bfx, y - \eta(\bfc,\lambda), 
       \cbfy + \bfc\Bigr)} 
       d\lambda = 0. \nonumber
\eeqn
for the $\tau$ function.  Notice here that the 
two factors in the denominator are power series 
of $\lambda^\ell$.  According to the second 
remark of the last theorem, one can insert any 
power series $f(\lambda^\ell)$ in the foregoing 
equation.  If we choose
\beqn
     f(\lambda^\ell) = \tau\Bigl(\bfx', y - \eta(\bfb,\lambda), 
       \cbfy + \bfb\Bigr)\tau\Bigl(\bfx, y - \eta(\bfc,\lambda), 
       \cbfy + \bfc\Bigr),
     \nonumber 
\eeqn
the two factors in the denominator are cancel out. 
The outcome is the bilinear equation 
\beqn
    \oint_{\lambda=\infty} && 
       \exp\Bigl(\xi(\bfx'-\bfx,\lambda)\Bigr) 
    \nonumber \\
    &\times&
       \tau\Bigl( \bfx' - \epsilon(\lambda), 
       y - \eta(\bfb,\lambda), 
       \cbfy + \bfb\Bigr) 
       \tau\Bigl(\bfx + \epsilon(\lambda), 
       y - \eta(\bfc,\lambda), 
       \cbfy + \bfc\Bigr) 
       d\lambda = 0. \nonumber
\eeqn
for the $\tau$ function. 


To rewrite this bilinear equation into the Hirota form, 
we replace $\bfx' \to \bfx + \bfa$, $\bfx \to \bfx - \bfa$, 
and choose $\bfc$ to be equal to $-\bfb$.  The bilinear 
equation can be thereby converted into a Hirota form: 
\beqn
    \oint_{\lambda=\infty} && 
    \exp\Bigl( 2\sum_{n=1}^\infty a_n \lambda^n \Bigr) 
    \exp\Bigl(-\sum_{n=1}^\infty\frac{D_{x_n}}{n}\lambda^{-n}\Bigr)
    \exp\Bigl(-\sum_{n=1}^\infty b_{n\ell}D_{y}\Bigr) 
    \nonumber \\
    &\times&
    e^{\langle\bfa,D_{\bfx}\rangle 
      + \langle\bfb,D_{\cbfy}\rangle}
    \tau(\bfx,\bfy)\cdot\tau(\bfx,\bfy) 
    d\lambda = 0. \nonumber
\eeqn
The exponential including $\lambda$ can be expanded 
into a series of Schur functions: 
\beqn
    \exp\Bigl(2\sum_{n=1}^\infty a_n\lambda^n\Bigr) 
    &=& \sum_{n=0}^\infty p_n(2\bfa) \lambda^n, 
    \nonumber \\
    \exp\Bigl(-\sum_{n=1}^\infty\frac{D_{x_n}}{n}\lambda^{-n}\Bigr) 
    &=& \sum_{n=0}^\infty p_n(-\widetilde{D_{\bfx}})\lambda^{-n}, 
    \nonumber \\
    \exp\Bigl(-\sum_{n=1}^\infty b_{n\ell}\lambda^{n\ell}
      D_{y}\Bigr)
    &=& \sum_{n=0}^\infty p_{n}(-\bfb D_{y})\lambda^{n\ell}. \nonumber
\eeqn
By plugging these identities into the last bilinear 
equations and do contour integrals, we obtain bilinear 
equations of the form 
\beqn
    \sum_{m,k=0}^\infty && 
    p_m(2\bfa) p_{m+k\ell+1}(-\widetilde{D_{\bfx}}) 
    p_{k}(-\bfb D_{y}) 
    \nonumber \\
    &\times& 
    e^{\langle\bfa,D_{\bfx}\rangle + \langle\bfb,D_{\cbfy}\rangle}
    \tau(\bfx,\bfy)\cdot\tau(\bfx,\bfy) 
    = 0.\nonumber 
\eeqn

These bilinear equations {\it almost} coincide with 
those that have been derived from the representation 
theory of toroidal algebras.  Although they do {\it not} 
agree, the discrepancy is rather superficial: 
If one shifts $\bfa$ to $\bfa + \kappa\,\bfb/2$ and 
move the extra exponential factor 
$\exp\Bigl(\kappa\sum_n b_{n\ell}\lambda^{n\ell}\Bigr)$ to 
the third exponential including $D_{y}$, 
the outcome exactly reproduces the previous result(cf. Remark
in the preceding subsection).

\section{Special solutions of hierarchy}\label{section:sol}
\setcounter{equation}{0}
\subsection{Construction of the solutions of the Wronskian type}
We shall apply the method developed by Date in 
(\cite{bib:Date1,bib:Date2,bib:Date3}) to construct special solutions of 
$\ell$-Bogoyavlensky hierarchy.
The method corresponds to consider
Baker-Akhiezer functions on 
rational singular curves with $N$ nodes.
Such algebro-geometric backgrounds
are given by Krichever in \cite{bib:Kr} (see also \cite{bib:DMN} of
Appendix 1 for an exposition)
and Manin in \cite{bib:Ma} Chap. III.
One can also consult instructive lectures
given by Mumford in \cite{bib:Mum}
(IIIb, section 5) that treat the subject.

As the data for the solution constructed below,
let us consider the following:
\begin{itemize}
\item $\alpha_i=\alpha_i(\bfy),\;c_i=c_i(\bfy) 
\;(i=1,\ldots,N)$
are arbitrary (local) solution of the equations
\beqn
\frac{\rd \alpha_i}{\rd y_{n\ell}}
-\alpha_i^{n\ell}
\frac{\rd \alpha_i}{\rd y}=0,\quad
\frac{\rd c_i}{\rd y_{n\ell}}
-\alpha_i^{n\ell}
\frac{\rd c_i}{\rd y}=0\quad n=1,2\ldots,\label{eq:RiemannWave}
\eeqn
such that we have $\alpha_i\not\equiv \alpha_j$ for
$i\ne j,$
\item $\zeta_i\;(i=1,\ldots,N)$ are $\ell$-th roots
of unity not equal to $1.$
\item $r_i\;(i=1,\ldots,N)$ are arbitrary
complex numbers.
\end{itemize}

We put $\beta_i(\bfy)\defeq\zeta_i\alpha_i(\bfy).$
Then we have the equations
\beqn
\frac{\rd \beta_i}{\rd y_{n\ell}}
-\beta_i^{n\ell}
\frac{\rd \beta_i}{\rd y}=0,\quad
n=1,2\ldots.\nonumber
\eeqn
Note that equation (\ref{eq:RiemannWave})
have constant solutions, namely, for distinct 
complex numbers $\alpha_i\;(i=1,\ldots,N)$
and arbitrary complex numbers $c_i\;(i=1,\ldots,N),$
$\alpha_i(\bfy)=\alpha_i,\;c_i(\bfy)=c_i$
trivially satisfy equations (\ref{eq:RiemannWave}).

Let us construct a special solution of
$\ell$-Bogoyavlensky hierarchy, which we shall
seek in the form 
\beqn
\Psi(\lambda,r)
=\left(
\lambda^N+w_1 \lambda^{N-1}+\cdots+w_N
\right)e^{\xi_r(\lambda)},\label{eq:WNPsi}
\eeqn
with $w_n=w_n(\bfx,\bfy)$ being unknown functions.
Here 
we set
$
\xi_r(\lambda)\defeq
\xi(\bfx,\lambda)+ry+r\eta(\cbfy,\lambda)
$
for any constant $r\in\bbC$ 
(cf. (\ref{eq:def_xi}),(\ref{eq:def_eta})).

We define the $N\times N$ matrix
\beqn
\Xi_N=\left(
\bff^{(0)},\bff^{(1)},\ldots,\bff^{(N-1)}
\right),\quad \bff^{(i)}\defeq\rd_x^i \bff\nonumber
\eeqn
where $\bff$ denote the $N$-column vector ${}^t(f_1,\ldots,f_N)$
and
\beqn
f_i\defeq
e^{\xi_{r_i}(\beta_i)}+c_i e^{\xi_{r_i}(\alpha_i)}.\label{eq:def_fi}
\eeqn 
Note that, for any constant solution $\{\alpha_i,c_i\}$
indicated above,
we clearly have $\det\Xi_N\not\equiv 0$.

\begin{theorem}
Let 
$
\{\alpha_i,\zeta_i,c_i,r_i\;(i=1,\ldots,N)
\}
$
be a given set of data described above. 
Suppose we have $\det\Xi_N\not\equiv 0$.
Then the condition
\beqn
\Psi(\beta_i,r_i)+c_i\Psi(\alpha_i,r_i)=0\quad 
(i=1,\ldots,N)\label{eq:singular}
\eeqn
uniquely determine the function $\Psi(\lambda,r)$ of the form
in (\ref{eq:WNPsi}), and 
$\Psi(\lambda)=\Psi(\lambda,r)$ solves equations (\ref{eq:lin_Psi_D}). 
Explicitly, we have the following expressions
\beqn
w_{N-k}=-\frac{\det\left(\bff^{(0)},\ldots,\bff^{(k-1)},
        \bff^{(N)},\bff^{(k+1)},\ldots,\bff^{(N-1)}\right)}
        {\det\left(\bff^{(0)},\ldots,\bff^{(k-1)},
        \bff^{(k)},\bff^{(k+1)},\ldots,\bff^{(N-1)}\right)}\quad
        (k=1,\ldots,N).
        \label{eq:Wron}
\eeqn
In particular, the corresponding $\tau$ function is given by 
the formula
\beqn
\tau=\det\Xi_N.\label{eq:tau_Wron}
\eeqn
\end{theorem}

\proof
Using the obvious differential equations
\beqn
\rd_{x}^n e^{\xi_r(\alpha_i)}=\alpha_i^n e^{\xi_r(\alpha_i)},
\quad 
\rd_{x}^n e^{\xi_r(\beta_i)}=\beta_i^n e^{\xi_r(\beta_i)},
\nonumber
\eeqn
the condition (\ref{eq:singular}) can be written in the
following form 
\beqn
W_N \bff=\bfzero\label{eq:singular-diff}
\eeqn
where by $W_N$ we denote the differential operator 
\beqn
W_N\defeq \rd_x^N+w_1\rd_x^{N-1}+\cdots+w_N.
\eeqn
Since $\det\Xi_N\not\equiv 0,$ 
we can solve equation 
(\ref{eq:singular-diff})
by Cramer's formula, 
to obtain expression
(\ref{eq:Wron}).

To show the function $\Psi(\lambda,r)$
satisfy (\ref{eq:linear}), we shall show 
that the PsDO defined by $W\defeq W_N\rd_x^{-N}$ 
solves equations (\ref{eq:Sato-x}) and (\ref{eq:Sato-y}).

It suffices to show equation (\ref{eq:Sato-x}) for {\textit{some}}
differential operator $B_n$ in $x$, since differential
operator $B_n$
is uniquely determined by the evolution equation itself.
Differentiating (\ref{eq:singular-diff}) with respect to
$x_n$ and using $\rd_{x_n}\bff=\rd_x^n\bff$, we have
\beqn
\left(\frac{\rd W_N}{\rd x_n}+
W_N{\rd_x^n}\right)\bff=\bfzero.\label{eq:WN-diff-xn}
\eeqn
There exist differential operators
$B_n$ and $R$  in $x$
such that 
\beqn
\frac{\rd W_N}{\rd x_n}+
W_N{\rd_x^n}=B_nW_N+R,\quad
\underset{\rd_x}{\ord}(R)\le N-1.\label{eq:WN_division}
\eeqn
>From (\ref{eq:singular-diff}) and (\ref{eq:WN-diff-xn})
we have
\beqn
R\bff=\bfzero\quad(i=1,\ldots,N).\nonumber
\eeqn
This implies $R=0$ since $\det\Xi_N\not\equiv 0$.
Hence by multiplying $\rd_x^{-N}$ from the right
to equation (\ref{eq:WN_division}) with $R=0,$
we obtain (\ref{eq:Sato-x}).

Now we recall $\alpha_i^\ell=\beta_i^\ell$. 
Since the dependence of
$f_i$ on $x_{n\ell}$
comes only through the overall factor
$\exp\left(\alpha_i^{n\ell}x_{n\ell}\right)
=\exp\left(\beta_i^{n\ell}x_{n\ell}\right),$
the expression (\ref{eq:Wron}) implies
\beqn
\frac{\rd W}{\rd x_{n\ell}}=0\quad (n=1,2\ldots).\nonumber
\eeqn
Hence from (\ref{eq:Sato-x})
$
W\rd_x^{n\ell}W^{-1}
$ is a differential operator, namely we have
\beqn
W\rd_x^{n\ell}W^{-1}=W_N\rd_x^{n\ell}W_N^{-1}=P^n,
\label{eq:def-Pn}
\eeqn
where $P$ denotes the differential operator 
$W\rd_x^\ell W^{-1}.$

Next we prove (\ref{eq:Sato-y}), the evolution
equation of $W$ with respect to $y_{n\ell}.$
Using (\ref{eq:RiemannWave}) and (\ref{eq:def_fi}) 
one has the equations
\beqn
\rd_{y_{n\ell}}\bff
=\left(\rd_x^{n\ell}\rd_y
-\Phi\right)\bff\quad \mbox{for}\quad n=1,2,\ldots,\nonumber\\
\mbox{where}\quad
\Phi\defeq\diag(\rd_y(\alpha_1^{n\ell}),\ldots,\rd_y(\alpha_N^{n\ell})).\nonumber
\eeqn
In view of these equations, the differentiation 
of (\ref{eq:singular-diff}) 
with respect to $y_{n\ell}$ yields
\beqn
\left(\frac{\rd W_N}{\rd y_{n\ell}}
+W_N\rd_x^{n\ell}\rd_y-W_N\Phi\right)\bff=\bfzero.
\label{eq:WN/y}
\eeqn
We use (\ref{eq:def-Pn}) to rewrite the operator 
$W_N\rd_y\rd_x^{n\ell}$ as follows
\beqn
W_N\rd_y\rd_x^{n\ell}
=W_N\rd_x^{n\ell}W_N^{-1}W_N\rd_y=P^nW_N\rd_y
=P^n\left(
-\frac{\rd W_N}{\rd y}+\rd_y\circ W_N
\right)\label{eq:PDOtoODO}
\eeqn
By virtue of (\ref{eq:PDOtoODO}), $W_N\Phi=\Phi W_N$ and 
(\ref{eq:singular-diff}), we 
can rewrite
(\ref{eq:WN/y}) as
\beqn
\left(\frac{\rd W_N}{\rd y_{n\ell}}
-P^n\frac{\rd W_N}{\rd y}\right)\bff=\bfzero.
\label{eq:WN-Sato-y}
\eeqn

Note that the operator in the left hand side of 
(\ref{eq:WN-Sato-y}) is an 
\textit{ordinary} differential operator,
which does not include $\rd_y.$
So we can apply exactly the same argument above 
to get the unique differential 
operator $C_{n\ell}$ in $x$ such that
\beqn
\frac{\rd W_N}{\rd y_{n\ell}}-P^n\frac{\rd W_N}{\rd y}
=C_{n\ell}W_N.\nonumber
\eeqn
This completes the proof of (\ref{eq:Sato-y}).
To see the formula (\ref{eq:tau_Wron}) is valid,
we only have to notice the following identity
arising from (\ref{eq:Psi&tau}) and (\ref{eq:WPsi}):
\beqn
w_1=-\rd_x\log\tau.
\eeqn
\qed
\subsection{$N$-soliton solutions}
If $\alpha_i,c_i(i=1,\ldots,N)$ in the preceding subsection
are constants, namely do not depend on $\bfy,$ then
the solution  of Wronskian type has clear 
representation-theoretical meaning.
In fact, we shall see that, 
up to an irrelevant factor,  
$\tau_N^{W}\defeq\det\Xi_N$ coincide 
with
the following $N${\textit{-soliton solution}}
\beqn
\tau_N^{S}\defeq
\left(1+a_N X(\alpha_N,\beta_N)V_{m_N}({\alpha_N^\ell})\right)
\cdots
\left(1+a_1 X(\alpha_1,\beta_1)V_{m_1}({\alpha_1^\ell})\right)
\vac_{\calB}^\tor,
\eeqn
where $a_i$ are constants we shall specify later.
To be rigorous, 
we have to assume $|\alpha_N|>\cdots>|\alpha_1|$
so that the series converges.
The fact that $\tau_N^{S}$ is a solution can be
readily observed in view of the next lemma 
(cf. Lemma \ref{lem:Omega-af}).
\begin{lemma}
Let $\zeta^\ell=1,\zeta\ne 1$ and $m\in \bbZ.$
Then on the representation 
$({\calF}_{\calB}^\tor,\pi_{\ell,\calB}^\tor)$
we have 
\beqn
\sum_{n\in\bbZ}E_n^\zeta t^m \lambda^{-n}
\mapsto
\frac{1}{1-\zeta}\Bigl(X(\lambda,\zeta\lambda)-1\Bigr)
V_m(\lambda^\ell).
\eeqn
\end{lemma}
\proof
Let us put
\beqn
      \Gamma(\lambda)\defeq
      \sum_{n\in \bbZ}E_n^\zeta t^m \cdot\lambda^{-n}
      +\frac{1}{1-\zeta}
      K^s_m(\lambda^\ell).
\eeqn
Then we have
\beqn
[\bphi_k,\Gamma(\lambda)]
&=&m\lambda^{k\ell}\Gamma(\lambda)\quad 
\forall k\in \bbZ\label{eq:bphi_Gamma}
\eeqn
and
\beqn
[\Lambda_k,\Gamma(\lambda)]
&=&(1-\zeta^k)\lambda^k\Gamma(\lambda)\quad 
\forall k\in \bbZ.\label{eq:Lambda_Gamma}
\eeqn
(\ref{eq:bphi_Gamma}) is a easy consequence of (\ref{eq:mult_s}),
and (\ref{eq:Lambda_Gamma}) follows from 
the following relation, which can be directly verified
\beqn
[\Lambda_k,E_n^\zeta t^m]
=
(1-\zeta^k)
\Bigl(
      E_{n+k}^\zeta t^m
      +\frac{\delta_{n+k,0}^{[\ell]}}{1-\zeta}
      \overline{s^{(n+k)/\ell}t^m d\log s}
\Bigr)\quad \forall k\in \bbZ
\eeqn
where $\delta^{[\ell]}_{m,n}= 1$ if $m\equiv n\mod \ell$
and $0$ otherwise.

By virtue of a standard lemma on vertex operators
(see e.g. \cite{bib:Kacbook} Lemma 14.5),
we have the conclusion that $\Gamma(\lambda)$ is realized on 
$\calF_{\calB}^\tor$ by 
\beqn
c_{\zeta,m}X(\lambda,\zeta\lambda)V_m(\lambda^\ell),
\quad c_{\zeta,m}\;\mbox{is a constant.}
\eeqn
To determine the constant $c_{\zeta,m}$,
we notice that
$
E_0^\zeta t^m\vac_{\calB}^\tor=0.
$
To see this, we only need
$E_0^\zeta \vac_{\calB}=0,$ which is obvious
from (\ref{eq:Ezeta_rep}), (\ref{eq:vac-psi}),
and the description of the representation in 
lemma \ref{lem:tor_ext}.
Now
it is immediate to see $c_{\zeta,m}=(1-\zeta)^{-1},$
and the proof completes.\qed

If we define the following matrices
\beqn
E_{\bfalpha}=\diag(e^{\xi(\alpha_1)},\ldots,
e^{\xi(\alpha_N)}),\quad
C=\diag(c_1,\ldots,c_N),\quad 
V_{\bfalpha}=(\alpha_i^{j-1})_{i,j=1,\ldots,N}.\nonumber
\eeqn
then
$
\Xi_N=E_{\bfbeta} V_{\bfbeta}+CE_{\bfalpha} V_{\bfalpha}
$ and hence we have
\beqn
\det\Xi_N=
\det E_{\bfbeta} \det V_{\bfbeta} \det\left(1+D\Gamma\right),\quad
D\defeq CE_{\bfalpha}E_{\bfbeta}^{-1},\quad
\Gamma\defeq V_{\bfalpha}V_{\bfbeta}^{-1}.\label{eq:detXi}
\eeqn

For any matrix $X=(x_{ij})_{i,j=1,\ldots,N}$ 
and any subset 
$J$ of $\{1,2,\ldots,N\}$, we define the principal submatrix
$X_J=(x_{jj'})_{j,j'\in J}.$ We shall use 
the well-known formula 
\beqn
\det(1+X)=\sum_{J\subset\{1,2,\ldots,N\}}\det X_J,
\quad J\;\mbox{runs over all the subsets of}\;\{1,2,\ldots,N\}.
\label{eq:Frob}
\eeqn

\begin{lemma}
Let $\Gamma\defeq V_{\bfalpha}V_{\bfbeta}^{-1}
=(\gamma_{ij})_{i,j\in I}.$
Then we have
\beqn
\gamma_{ij}=\displaystyle
\frac{\prod_{k(\ne j)}(\alpha_i-\beta_k)}
{\prod_{k(\ne j)}(\beta_j-\beta_k)}\quad\mbox{and}\quad
\frac{\det(\Gamma_J)}{\prod_{j\in J}\gamma_{jj}}
=\prod_{j,j'\in J,j<j'}\frac{(\alpha_j-\alpha_{j'})(\beta_j-\beta_{j'})}
{(\alpha_j-\beta_{j'})(\beta_j-\alpha_{j'})}\quad
\mbox{for}\;\forall J\subset I.\nonumber
\eeqn
\end{lemma}
Combining the formulas (\ref{eq:detXi}),(\ref{eq:Frob}), and 
the above lemma (cf. \cite{bib:Mum}), we have
\beqn
\tau_N^W=
e^{\sum_{i=1}^N\xi_{m_i}(\beta_i)}
\Delta(\bfbeta)
\sum_{J\subset I}
\left(
\prod_{j\in J}c_j\gamma_{jj}
\right)
\underset{j<j'}{\underset{j,j'\in J}{\prod}}
\frac{(\alpha_j-\alpha_{j'})(\beta_j-\beta_{j'})}
{(\alpha_j-\beta_{j'})(\beta_j-\alpha_{j'})}
e^{\sum_{j\in J}(\xi_{m_j}(\alpha_j)-\xi_{m_j}(\beta_j))},\nonumber
\eeqn
where $\Delta(\bfbeta)$ denotes Vandermonde's determinant
$\det V_{\bfbeta}=\prod_{i>j}(\beta_i-\beta_j).$
Here we also note that $\tau_N^S$ does not 
depend on $y^*_{n\ell}$ from the construction.
So if we set $a_i=c_j\gamma_{jj},$ 
we can see that 
$\tau_N^{W}$ is equal to $\tau_N^S$
times an irrelevant factor 
$e^{\sum_{i=1}^N\xi_{m_i}(\beta_i)}
\Delta(\bfbeta).$

\section{Conclusion}

Inspired by the work of Billig and Iohara et al.,
we have introduced the $\ell$-Bogoyavlensky hierarchy
and clarified its group-theoretic properties, Lax
formalism, and special solutions.  The group-theoretic
characterization of the new hierarchy is based on
the $2$-toroidal Lie algebra ${\fraksl}_\ell^{\tor}$.
We have constructed a representation of this Lie
algebra, and derived a system of Hirota bilinear
equations for the $\tau$ function.  The point of
departure of the Lax formalism is the observation
that the lowest two members of the Hirota bilinear
equations of Billig and Iohara et al. are equivalent
to Bogoyavlensky's $2+1$ dimensional extension of
the KdV equation.  Bearing this in mind, we have
constructed  a system of Lax equations, from which
we have been able to reproduce the Hirota bilinear
equations.  The Lax formalism has also turned out
to be useful for understanding special solutions.

Several interesting problems remain open.  Firstly,
the present construction should be extended to
the toroidal Lie algebras associated with simple
Lie algebras other than $\fraksl_\ell$.  This
will give an extension of the Drinfeld-Sokolov
hierarchies \cite{bib:DS} in the spirit of Bogoyavlensky.
The same problem can also be raised to another
important family of soliton equations, namely
those represented by the nonlinear Schr\"odinger
equation.  Strachan's work \cite{bib:St} is
remarkable in this respect. He presented a
$2+1$-dimensional extension and an associated
hierarchy of higher equations for soliton
equations of this type.  Lastly, we would
like to mention the problem of constructing
special solutions of the algebro-geometric
(``finite-band'') type.  A precursor of
this problem is Cherednik's work \cite{bib:Ch},
in which he presented a construction of
algebro-geometric solutions to the self-dual
Yang-Mills equation.  Since our hierarchy is
closely related to the self-dual Yang-Mills
equation, a similar construction of special
solutions will be possible.

\section{Appendix}\label{section:app}
{ We give a list of the Hirota equation of low degree
contained in the hierarchy for $\ell=2$. Since we have $P(D)f\cdot f=0$
for any function $f$ and polynomial $P$ such that $P(-D)=-P(D)$, we
shall list below only the even polynomials in $D$. We also drop the terms 
including $D_{x_2},D_{x_4},\ldots$. We shall use the 
abbreviated notation $D_n$ and $Q_n$ for
$D_{x_n}$ and $D_{y_n}$ respectively.
 
\begin{center}
   \begin{tabular}{|l|l|}                         \hline
                   degree 3  & $D_1^3 Q_0 + 2 D_3 Q_0 - 6 D_1 Q_2$ \\ \hline
                   degree 4  & $D_1^4-4D_1 D_4$                    \\ \hline
                   degree 5  & $\left(D_1^5
                                 +20 D_1^2 D_3
                                 +24 D_5 
                                 \right) Q_0
                                 -120 D_1 Q_4$                     \\ \cline{2-2}
                             & $\left(
                                 D_1^5
                                 +20 D_1^2 D_3
                                 +24 D_5  
                                 \right) Q_0
                                 -\left(
                                 20 D_1^3 
                                 +40 D_3 
                                 \right) Q_2$                      \\ \cline{2-2}
                             & $\left(
                                 D_1^5
                                 -10 D_1^2 D_3
                                 +24 D_5
                                 \right) Q_0
                                 +\left(
                                 10 D_1^3
                                 -40 D_3 
                                 \right) Q_2$                      \\ \hline
                    degree 6  & $\left(
                                 D_1^6
                                 -20 D_3^2
                                 +10 D_1^3 D_3
                                 -36 D_1 D_5 
                                 \right) Q_0^2 
                                 -30\left(
                                 D_1^4 
                                 -4 D_3 D_1
                                 \right) Q_0 Q_2$                  \\ \cline{2-2}
                              &  $D_1^6
                                 -32 D_3^2
                                 +4 D_1^3 D_3$                     \\ \cline{2-2}
                              &  $D_1^6 
                                 -80 D_3^2 
                                 -20 D_1^3 D_3
                                 +144 D_1 D_5$                     \\ \hline
                     degree 7 &  $\left( 
                                 D_1^7 
                                 + 720 D_7   
                                 + 70 D_1^4 D_3 
                                 + 280 D_3^2 D_1
                                 + 504 D_5 D_1^2 
                                 \right) Q_0                 
                                 $ \\
                              &  $-\left(
                                 420 D_1^2 D_3 
                                 + 21 D_1^5
                                 + 504 D_5
                                 \right) Q_2- \left(
                                 420 D_1^3 
                                 + 840 D_3 
                                 \right) Q_4$                      \\ \cline{2-2}
                              &  $\left(
                                 3 D_1^7
                                 - 168 D_5 D_1^2 
                                 + 480 D_7
                                 \right) Q_0 
                                 -\left(
                                 1120 D_3 
                                 -280 D_1^3 
                                 \right) Q_4$                      \\ \cline{2-2}
                              &  $\left(
                                 3 D_1^7
                                 - 168 D_5 D_1^2 
                                 + 480 D_7 
                                 \right) Q_0
                                 + \left(
                                 280 D_1^2 D_3 
                                 - 28 D_1^5 
                                 - 672 D_5 
                                 \right) Q_2$                      \\ \cline{2-2}
                              &  $\left(
                                 5 D_1^7 
                                 - 1440 D_7 
                                 - 70 D_1^4 D_3 
                                 + 560 D_3^2 D_1 
                                 \right) Q_0 
                                 -\left(
                                 126 D_1^5 
                                 - 2016 D_5 
                                 \right) Q_2$                      \\ \cline{2-2}
                              &  $\left(
                                 D_1^7 
                                 + 720 D_7
                                 + 70 D_1^4 D_3
                                 + 280 D_3^2 D_1
                                 + 504 D_5 D_1^2 
                                 \right) Q_0$ \\
                              &  $-\left(
                                 840 D_1^2 D_3 
                                 + 42 D_1^5 
                                 + 1008 D_5
                                 \right) Q_2$                      \\ \cline{2-2}  
                              &  $\left(
                                 D_1^7  
                                 + 720 D_7
                                 + 70 D_1^4 D_3 
                                 + 280 D_3^2 D_1 
                                 + 504 D_5 D_1^2 
                                 \right) Q_0
                                 - 5040 D_1 Q_6$                   \\ \cline{2-2}
                              &  $\left(
                                 D_1^7
                                 + 720 D_7
                                 + 70 D_1^4 D_3
                                 + 280 D_3^2 D_1
                                 + 504 D_5 D_1^2  
                                 \right) Q_0^3 - 5040 D_1 Q_2^3 $          \\
                              &  $-\left( 
                                 2520 D_1^2 D_3 
                                 + 126 D_1^5
                                 + 3024 D_5 
                                 \right) Q_0^2 Q_2 
                                 +\left(
                                 2520 D_1^3
                                 +5040 D_3
                                 \right) Q_0 Q_2^2$                        \\ \hline                          
    \end{tabular}
\end{center}

\subsection*{Acknowledgements}

We are grateful to Koji Hasegawa, Kenji Iohara, 
Tetsuya Kikuchi, Gen Kuroki, 
Takashi Takebe,
Yoshihisa Saito,  Ryuichi Sawae, and Takahiro Shiota
for fruitful discussions.
This work was partly supported by the 
Grant-in-Aid for Scientific Research (No. 10640165) 
from the Ministry of Education, Science and Culture.

\end{document}